\documentclass[
  journal=pasa,
  manuscript=research-paper, %% or "review"
  year=2022,
  volume=37,
]{cup-journal}

\usepackage{microtype,siunitx,booktabs}
\usepackage{amsmath}
\usepackage{amssymb}
\usepackage{multirow}
\usepackage{threeparttable}
\usepackage[figuresright]{rotating}
\usepackage{subcaption}
\usepackage{hyperref}
\usepackage{float}

\hypersetup{colorlinks=true,citecolor=blue,linkcolor=blue,urlcolor=blue}

\newcommand{\pinta}{\texttt{pinta}}

\newcommand{\rficlean}{\texttt{RFIClean}}

\newcommand{\dspsr}{\texttt{dspsr}} 

\newcommand{\psrchive}{\texttt{PSRCHIVE}}
\newcommand{\psrfits}{\texttt{PSRFITS}}
\newcommand{\tempo}{\texttt{TEMPO}}
\newcommand{\tempotwo}{\texttt{TEMPO2}}
\newcommand{\dmcalc}{\texttt{DMcalc}}
\newcommand{\pulseportr}{\texttt{PulsePortraiture}}

\title[InPTA DR1]{The Indian Pulsar Timing Array: First data release}

\author{Pratik Tarafdar}
\affiliation{The Institute of Mathematical Sciences, CIT Campus, Taramani, Chennai 600113, Tamil Nadu, India}
\email[Pratik Tarafdar]{pratikta16@gmail.com}

\author{Nobleson K.}
\affiliation{Department of Physics, BITS Pilani Hyderabad Campus, Hyderabad 500078, Telangana, India}

\author{Prerna Rana}
\affiliation{Department of Astronomy and Astrophysics, Tata Institute of Fundamental Research, Homi Bhabha Road, Navy Nagar, Colaba, Mumbai 400005, INDIA}

\author{Jaikhomba Singha}
\affiliation{Department of Physics, Indian Institute of Technology Roorkee, Roorkee 247667, Uttarakhand, India}

\author{M. A. Krishnakumar}
\affiliation{Max-Planck-Institut f{\"u}r Radioastronomie, Auf dem H{\"u}gel 69, 53121 Bonn, Germany}
\alsoaffiliation{Fakult{\"a}t f{\"u}r Physik, Universit{\"a}t Bielefeld, Postfach 100131, 33501 Bielefeld, Germany}

\author{Bhal Chandra Joshi}
\affiliation{National Centre for Radio Astrophysics, Pune University Campus, Pune 411007, India}

\author{Avinash Kumar Paladi}
\affiliation{Department of Earth \& Space Sciences, IIST Thiruvananthapuram, Kerala 695547, India}

\author{Neel Kolhe}
\affiliation{Department of Physics, St. Xavier’s College (Autonomous), Mumbai 400001, Maharashtra, India}

\author{Neelam Dhanda Batra}
\affiliation{Department of Physics \& Astrophysics, University of Delhi, Delhi-110007}

\author{Nikita Agarwal}
\affiliation{Department of Electronics and Communication Engineering, Manipal Institute of Technology, Manipal Academy of Higher Education, Manipal 576104, Karnataka, India}

\author{Adarsh Bathula}
\affiliation{Department of Physics, Indian Institute of Science Education and Research, Mohali, Punjab 140306}

\author{Subhajit Dandapat}
\affiliation{Department of Astronomy and Astrophysics, Tata Institute of Fundamental Research, Homi Bhabha Road, Navy Nagar, Colaba, Mumbai 400005, INDIA}

\author{Shantanu Desai}
\affiliation{Department of Physics, IIT Hyderabad, Kandi, Telangana 502284, India}

\author{Lankeswar Dey}
\affiliation{National Centre for Radio Astrophysics, Pune University Campus, Pune 411007, India}
\alsoaffiliation{Department of Astronomy and Astrophysics, Tata Institute of Fundamental Research, Homi Bhabha Road, Navy Nagar, Colaba, Mumbai 400005, INDIA}

\author{Shinnosuke Hisano}
\affiliation{Kumamoto University, Graduate School of Science and Technology, Kumamoto, 860-8555, Japan}

\author{Prathamesh Ingale}
\affiliation{Fergusson College (Autonomous), Pune-04,Maharashtra, India}

\author{Ryo Kato}
\affiliation{Osaka Central Advanced Mathematical Institute, Osaka Metropolitan University, Osaka, 5588585, Japan}
\alsoaffiliation{Faculty of Advanced Science and Technology, Kumamoto University, 2-39-1 Kurokami, Kumamoto 860-8555, Japan}

\author{Divyansh Kharbanda}
\affiliation{Department of Physics, IIT Hyderabad, Kandi, Telangana 502284, India}

\author{Tomonosuke Kikunaga}
\affiliation{Kumamoto University, Graduate School of Science and Technology, Kumamoto, 860-8555, Japan}

\author{Piyush Marmat}
\affiliation{Department of Physics, Indian Institute of Technology Roorkee, Roorkee 247667, Uttarakhand, India}

\author{B. Arul Pandian}
\affiliation{Raman Research Institute, Bengaluru, Karnataka, India.}

\author{T. Prabu}
\affiliation{Raman Research Institute, Bengaluru, Karnataka, India.}

\author{Aman Srivastava}
\affiliation{Department of Physics, IIT Hyderabad, Kandi, Telangana 502284, India}

\author{Mayuresh Surnis}
\affiliation{Jodrell Bank Centre for Astrophysics, Department of Physics and Astronomy, University of Manchester, Manchester M13 9PL, UK}

\author{Sai Chaitanya Susarla}
\affiliation{School of Mathematics, National University of Ireland Galway, University road, Galway, H91TK33, Ireland}

\author{Abhimanyu Susobhanan}
\affiliation{National Centre for Radio Astrophysics, Pune University Campus, Pune 411007, India}
\alsoaffiliation{Center for Gravitation, Cosmology, and Astrophysics, University of Wisconsin-Milwaukee, Wisconsin 53211, USA}

\author{Keitaro Takahashi}
\affiliation{Faculty of Advanced Science and Technology, Kumamoto University, 2-39-1 Kurokami, Kumamoto 860-8555, Japan}
\alsoaffiliation{International Research Organization for Advanced Science and Technology, Kumamoto University, 2-39-1 Kurokami, Kumamoto 860-8555, Japan}

\author{P. Arumugam}
\affiliation{Department of Physics, Indian Institute of Technology Roorkee, Roorkee 247667, Uttarakhand, India}

\author{Manjari Bagchi}
\affiliation{The Institute of Mathetical Sciences, CIT Campus, Taramani, Chennai 600113, Tamil Nadu, India}
\alsoaffiliation{Homi Bhabha National Institute, Training School Complex, Anushakti Nagar, Mumbai 400094, India}

\author{Sarmistha Banik}
\affiliation{Department of Physics, BITS Pilani Hyderabad Campus, Hyderabad 500078, Telangana, India}

\author{Kishalay De}
\affiliation{MIT Kavli Institute for Astrophysics and Space Research, 77 Massachusetts Ave, Cambridge, MA 02139, USA}

\author{Raghav Girgaonkar}
\affiliation{Amity Centre of Excellence in Astrobiology, Amity University Mumbai 410206, Maharashtra, India}

\author{A. Gopakumar}
\affiliation{Department of Astronomy and Astrophysics, Tata Institute of Fundamental Research, Homi Bhabha Road, Navy Nagar, Colaba, Mumbai 400005, INDIA}

\author{Yashwant Gupta}
\affiliation{National Centre for Radio Astrophysics, Pune University Campus, Pune 411007, India}

\author{Yogesh Maan}
\affiliation{National Centre for Radio Astrophysics, Pune University Campus, Pune 411007, India}

\author{P. K. Manoharan}
\affiliation{Arecibo Observatory, University of Central Florida, Arecibo 00612, USA}

\author{Arun Naidu}
\affiliation{University of Oxford, Denys Wilkinson Building, Keble Road, Oxford, UK, OX13RH}

\author{Dhruv Pathak}
\affiliation{The Inter-University Centre for Astronomy and Astrophysics, Pune-411007, India}

\doi{xx.yyyy/pasa.zzzz.tt}

\received {dd Mmm YYYY}
\revised  {dd Mmm YYYY}
\accepted {dd Mmm YYYY}
\published{dd Mmm YYYY}

\keywords{radio telescopes; radio astronomy; astronomy data analysis; pulsar timing method; millisecond pulsars}

\begin{document}

\begin{abstract} 
We present the pulse arrival times and high-precision dispersion measure estimates for 14 millisecond pulsars observed simultaneously in the 300$-$500 MHz and 1260$-$1460 MHz frequency bands using the upgraded Giant Metrewave Radio Telescope (uGMRT). The data spans over a baseline of 3.5 years (2018-2021), and is the first official data release made available by the Indian Pulsar Timing Array collaboration. This data release presents a unique opportunity for investigating the interstellar medium effects at low radio frequencies and their impact on the timing precision of pulsar timing array experiments. In addition to the dispersion measure time series and pulse arrival times obtained using both narrowband and wideband timing techniques, we also present the dispersion measure structure function analysis for selected pulsars. Our ongoing investigations regarding the frequency dependence of dispersion measures have been discussed. Based on the preliminary analysis for five millisecond pulsars, we do not find any conclusive evidence of chromaticity in dispersion measures. Data from regular simultaneous two-frequency observations are presented for the first time in this work. This distinctive feature leads us to the highest precision dispersion measure estimates obtained so far for a subset of our sample. Simultaneous multi-band uGMRT observations in Band 3 and Band 5 are crucial for high-precision dispersion measure estimation and for the prospect of expanding the overall frequency coverage upon the combination of data from the various Pulsar Timing Array consortia in the near future. Parts of the data presented in this work are expected to be incorporated into the upcoming third data release of the International Pulsar Timing Array.
\end{abstract}

\section{Introduction}
\label{sec:intro}

Pulsars are rotating neutron stars whose electromagnetic radiation is received as periodic pulses by the observers. 
Their high rotational stability allows us to use them as accurate celestial clocks to probe a wide range of time-domain phenomena \citep{hgc+2019}.
This is possible through the technique of pulsar timing, where the pulsar rotation is accurately tracked by precisely measuring the times of arrival (ToAs) of its pulses \citep{ehm2006}. 
Pulsar Timing Arrays \citep[PTAs:][]{fb1990} such as the Parkes Pulsar Timing Array \citep[PPTA:][]{h2013,mhb+2013,krh+2020}, the Indian Pulsar Timing Array \citep[InPTA:][]{jab+2018}, the European Pulsar Timing Array \citep[EPTA:][]{kc2013,dcl+2016,ccg+2021}, and the North American Nanohertz Observatory for Gravitational Waves \citep[NANOGrav:][]{m2013,dfg+2013,abb+2015,abb+2018a,aab+2020a,aab+2020b} are experiments that aim to detect gravitational waves (GWs) in the nanohertz frequency range using an ensemble of millisecond pulsars (MSPs) as a galaxy-sized celestial detector.
The International Pulsar Timing Array \citep[IPTA:][]{haa+2010,vlh+2016,pdd+2019} aims to combine the data and resources from the different PTA experiments to expedite the detection of nanohertz GWs and strengthen the  post-detection science capabilities.
These efforts, together with rapidly maturing PTA campaigns in China \citep{l2016} and South Africa \citep{bbb+2018},  are expected to inaugurate the field of nanohertz GW astronomy in the near future \citep{gsr+2021,abb+2021,ccg+2021}.

The InPTA experiment aims to use the unique strengths of the Giant Metrewave Radio Telescope \citep[GMRT:][]{sak+1991} after its recent major upgrade \citep[uGMRT:][]{gak+2017} to complement the international PTA efforts by providing a unique low-frequency view of PTA pulsars.
In our experiment, the data are recorded \textit{simultaneously} in both 300$-$500 MHz and 1260$-$1460 MHz ranges, which enable us to make some of the most precise dispersion measure estimations so far. While there are  other low-frequency telescopes, such as the Murchison Widefield Array (70$-$300 MHz) and LOFAR (10$-$240 MHz), the distinctive feature of the uGMRT is its ability to observe simultaneously over a much wider frequency range including low frequencies with multiple bands. It may be noted that similar capability is likely in the future with FAST with its wideband receivers (70 MHz$-$3 GHz) and Square Kilometer Array, but currently, only the uGMRT has such capabilities.
Such low-frequency observations are especially relevant for characterizing the effects of the interstellar medium, which are strongest at low frequencies \citep{lk2004}.
The broad bandwidth and high collecting area provided by the uGMRT at low radio frequencies make it an ideal instrument to characterize such effects \citep{kmj+2021}.

We describe in this paper the time of arrival (ToA) measurements, timing analysis, and the characterization of dispersion measure variations of 14 pulsars using observations taken over a span of 3.5 years (2018$-$2021) with the uGMRT.
In this work, we employ two different approaches for ToA generation and dispersion measure (DM) estimation. 
The traditional `narrowband' approach splits the broadband observations into multiple sub-bands. 
The ToAs are computed separately for each sub-band \citep{t1992}, and the DM for each epoch can be estimated from these narrowband ToAs.
We refer to these measurements as the narrowband ToAs and the narrowband DMs.
An alternative approach, employing the more recent `wideband' technique \citep{pdr2014,p2015} utilises principal component decomposition of frequency-resolved pulsar profiles observed over a wide bandwidth to account for the frequency-dependent evolution of the profiles. 
This technique generates a single ToA per observation for the entire bandwidth, together with the corresponding DM.
We refer to the ToAs and DMs estimated using this method as the wideband ToAs and wideband DMs, respectively.
Note that the wideband DMs are not derived quantities of the wideband ToAs unlike in the narrowband case.
\citet{nag+2022} showed that the application of the wideband technique  to InPTA data provides significant improvement in the timing precision as compared to the traditional narrowband technique.

The data presented here, particularly the high precision DMs, are likely to enhance the sensitivity required to detect nanohertz GWs when these are combined with data from other PTA experiments. Simulations available in the literature suggest that accommodating epoch-wise high precision DM measurements  can improve the precision of the estimated pulsar parameters as well as the upper limit/estimate of the amplitude of red noise in a PTA dataset significantly \citep{lah13arxiv,pdr2014,lsc16mnras,grs21mnras,cbp22mnras}.

The data presented in this paper is being made available publicly for combinations with other PTA experiments\footnote{\url{https://github.com/inpta/InPTA.DR1}} (refer Section \ref{sec:data}). 
Parts of the data presented in this work are expected to be included in the upcoming third data release of the International Pulsar Timing Array.

This article is arranged in the following manner. 
A brief overview of the InPTA experiment is provided in Section \ref{sec:inpta}. 
In Section \ref{sec:obs}, we describe our observation, data reduction, and pre-processing procedures.
In Section \ref{sec:toas_dms}, we describe the methods for estimating ToAs and DMs from our observations.
Section \ref{sec:narrowband} elaborates our iterative narrowband method of ToA generation and DM estimation.
Section \ref{sec:wide-band} explains the methodology adopted for ToA and DM estimation using the wideband technique.
Section \ref{sec:timing} consists of details regarding the procedure of obtaining narrowband and wideband timing solutions.
Section \ref{sec:sfmethod} includes a description of the methodology used for structure function analysis of the obtained DM time series.
The estimated ToAs, DM variations, the timing model used for each pulsar, the estimated timing parameters, results from DM structure function analysis, and preliminary investigations into the frequency-dependence of DMs are presented in Section \ref{sec:result}.
Finally, in Section \ref{sec:conclusions}, we summarize our work and discuss future directions.

\section{The InPTA Experiment}  
\label{sec:inpta}

\begin{figure*}
\centering
\includegraphics[width=0.7\linewidth]{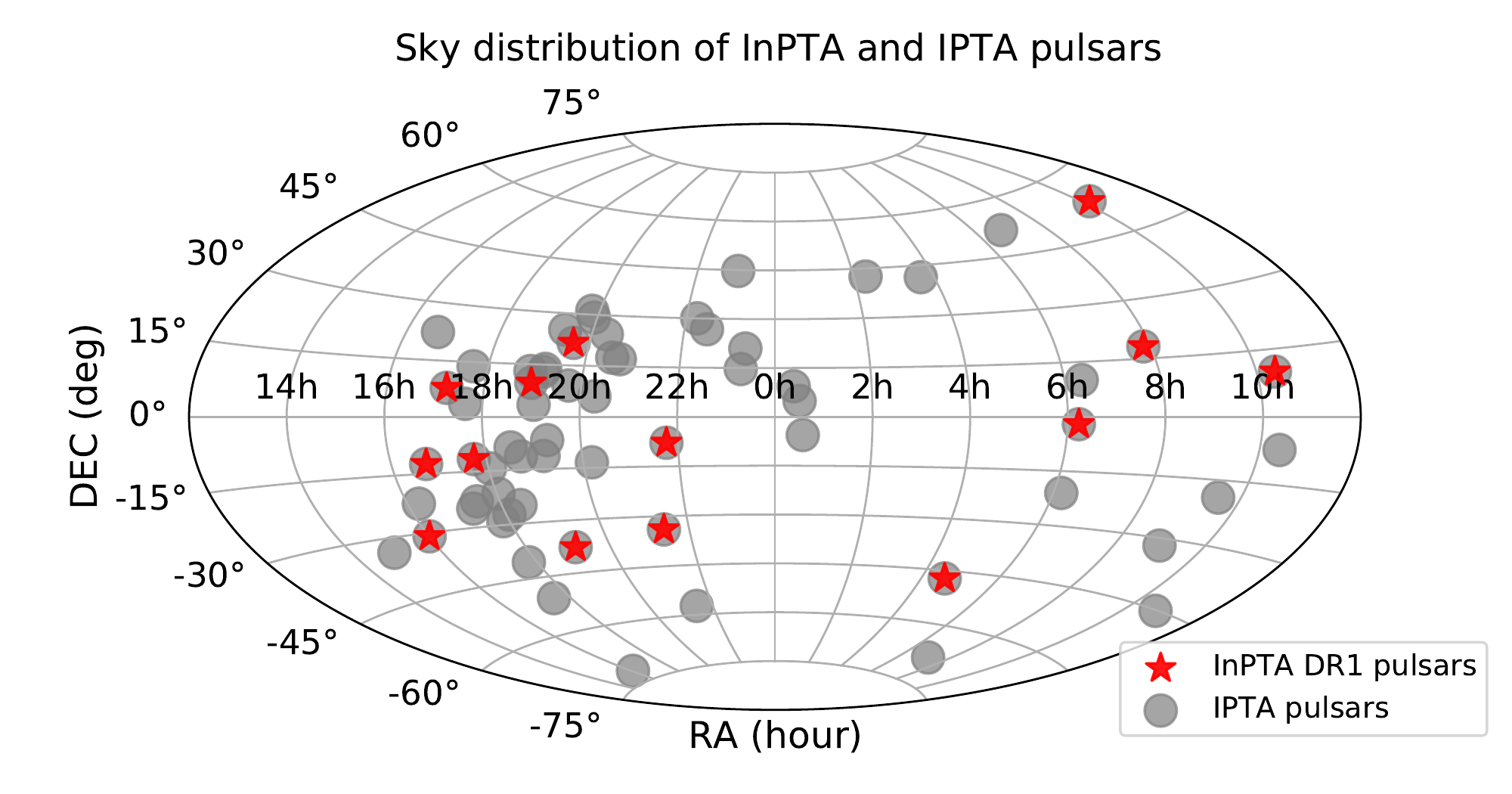}
\caption{The sky distribution of those 14 pulsars observed during the InPTA experiment that are included in the present data release is indicated by red stars, whereas that used in the IPTA Data Release 2 is marked with grey circles. }
\label{fig:galdist}
\end{figure*}

The Indian Pulsar Timing Array experiment\footnote{\url{https://inpta.iitr.ac.in/}} started as a pilot campaign using the legacy GMRT and the Ooty Radio Telescope \citep[ORT:][]{ssj+1971} in 2015.
The experiment is conducted by a multi-national collaboration of about 40 researchers hailing from various institutions. 
During this pilot phase of the experiment, we tested our observation and data analysis strategies, defined the pulsar sample, and obtained preliminary timing solutions for folding the pulsar time-series data.
The InPTA started using the new capabilities of the upgraded GMRT since April 2018. 
The initial sample for the experiment consisted of 22 pulsars which were observed once every 15 days. 
As the collaboration experimented with different observing strategies, the number of observed pulsars ranged from 5 to 22.
A discussion of these limitations during the initial phases of the experiment is beyond the scope of the present article. All such observation strategies are elaborated in \cite{jgp+2022}.
The sky distribution of the 14 selected pulsars from the full InPTA sample that are included in the present data release is shown in Figure \ref{fig:galdist} in comparison with the sample used in the International Pulsar Timing Array Data Release 2 \citep[IPTA DR2:][]{pdd+2019}. 
Except for a hiatus of 6 months during 2019, at least 5 pulsars have been routinely observed for the last three and a half years using the uGMRT since the beginning of the main InPTA experiment. 
The first data release of the InPTA, consisting of data products for fourteen pulsars, will be made publicly available together with this paper.
The details of the InPTA observations and data analysis are presented in the forthcoming sections.

\section{Observations and data processing}
\label{sec:obs}

In this work, we use observations of 14 pulsars conducted using the uGMRT \citep{gak+2017} as part of the InPTA experiment from 2018 to 2021 typically with a bi-weekly cadence. 
These observations were carried out during observing cycles 34$-$35 and 37$-$40 of the uGMRT, where the 30 uGMRT antennae were divided into multiple phased subarrays, simultaneously observing the same source in multiple bands in total intensity mode.
During cycles 34$-$35, the observations were carried out using three subarrays in Band 3 (400$-$500 MHz), Band 4 (650$-$750 MHz), and Band 5 (1360$-$1460 MHz) of the uGMRT with 100 MHz bandwidth, whereas in cycles 37$-$40, the observations were carried out using two subarrays in Band 3 (300$-$500 MHz) and Band 5 (1260$-$1460 MHz) with 200 MHz bandwidth.
As observations in Band 4 were carried out only in the earliest two cycles with 100 MHz of bandwidth, this dataset is not included in the present data release.
The Band 3 data in all cycles as well as the Band 5 data in cycles 34$-$35 (except observations between Oct. 20, 2018 and Nov. 14, 2018) were coherently dedispersed using a real-time pipeline \citep{dg2016} to the known DM of each pulsar.
The setup and command files for the observation sessions are prepared using automated scripts developed and maintained by the InPTA members.
The observation settings used in general for different observation cycles, such as the number of frequency channels and the sampling time, are listed in Table \ref{tab:observations} (non-standard epochs are mentioned in the caption).
The sky distribution of the pulsars observed between 2018$-$2021 is plotted in Figure \ref{fig:galdist}.

\begin{table*}
\begin{tabular}{|c|c|c|c|c|c|c|c|c|}
\hline
\hline
\textbf{\begin{tabular}[c]{@{}c@{}}Observation\\Cycle\end{tabular}} &
\textbf{\begin{tabular}[c]{@{}c@{}}No. of\\PSRs\end{tabular}} &
\textbf{\begin{tabular}[c]{@{}c@{}}MJD \\ Start\end{tabular}} & \textbf{\begin{tabular}[c]{@{}c@{}}MJD\\ End\end{tabular}} & \textbf{\begin{tabular}[c]{@{}c@{}}Band\\ No\end{tabular}} & \textbf{\begin{tabular}[c]{@{}c@{}}Frequency\\ Band (MHz)\end{tabular}} & \textbf{\begin{tabular}[c]{@{}c@{}}Number of \\ channels\end{tabular}} & \textbf{\begin{tabular}[c]{@{}c@{}}Sampling\\ time ($\mu$s)\end{tabular}} & \textbf{\begin{tabular}[c]{@{}c@{}}Coherent\\ Dedispersion\end{tabular}} \\ \hline
                             & 22 &    58235                   &    58389                      & 3   & 400$-$500     & 1024   & 81.92  & Yes   \\ \cline{5-9} 
                            &                       & &                          & 4   & 650$-$750     & 1024   & 81.92  & No    \\ \cline{5-9} 
\multirow{-3}{*}{34}        & \multirow{-3}{*}{}    & \multirow{-3}{*}{}       & & 5   & 1360$-$1460   & 1024   & 81.92  & Yes   \\ \hline
                             & 22 &  58413                     &      58524                    & 3   & 400$-$500     & 1024   & 81.92  & Yes   \\ \cline{5-9} 
                            &                       & &                          & 4   & 650$-$750     & 1024   & 81.92  & No    \\ \cline{5-9} 
\multirow{-3}{*}{35}        & \multirow{-3}{*}{}    & \multirow{-3}{*}{}       & & 5   & 1360$-$1460   & 1024   & 81.92  & Yes   \\ \hline
                             & 6 &  58781                     &     58922                     & 3   & 300$-$500     & 512    & 20.48  & Yes   \\ \cline{5-9} 
\multirow{-2}{*}{37}        & \multirow{-2}{*}{}    & \multirow{-2}{*}{}       & & 5   & 1260$-$1460   & 1024   & 40.96  & No    \\ \hline
                             & 5 &    58990                   &    59133                      & 3   & 300$-$500     &   512     & 20.48       & Yes   \\ \cline{5-9} 
\multirow{-2}{*}{38}        & \multirow{-2}{*}{}    & \multirow{-2}{*}{}       & & 5   & 1260$-$1460   &    1024    &   40.96     & No    \\ \hline
                             & 6 &  59156                     &      59309                    & 3   & 300$-$500     &    256    &  10.24      & Yes   \\ \cline{5-9} 
\multirow{-2}{*}{39}        & \multirow{-2}{*}{}    & \multirow{-2}{*}{}       & & 5   & 1260$-$1460   &    1024    &   40.96     & No    \\ \hline
                             & 13 &   59343                    &     59496                     & 3   & 300$-$500     &    128    &    5.12    & Yes   \\ \cline{5-9} 
\multirow{-2}{*}{40}     & \multirow{-2}{*}{}    & \multirow{-2}{*}{}       & & 5   & 1260$-$1460   &    1024    &   40.96     & No    \\ \hline
%                            &   59516                    &     59672                     & 3   & 300-500     &    128    &    5.12    & Yes   \\ \cline{4-8} 
%\multirow{-2}{*}{41}     & \multirow{-2}{*}{}    & \multirow{-2}{*}{}       & 5   & 1260-1460   &    1024    &   40.96     & No    \\ \hline
\end{tabular}
\caption{Observation settings used for InPTA observations. 
The multi-band observations are carried out simultaneously using multiple sub-arrays, and recorded using the GWB backend. A +10kHz correction is applied for MJDs lying between 59217 and 59424 (Cycle 39$-$40) due to an offset in the local oscillator frequencies in all bands at the observatory during these epochs. PSRs were observed on MJDs 58413 and 58431 (Cycle 35) with a bandwidth of 200 MHz (standard bandwidth for cycles 34 and 35 was 100 MHz). Band 5 data between MJDs 58411 and 58436 (Cycle 35) were not recorded with coherent dedispersion. MJDs 59376 and 59380 (Cycle 40) were non-standard observations with the Polyphase Filterbank (PFB) setting turned on (PFB is turned off for all our observations in general).}
%\warn{AS}{Please mention the non-standard observation dates present in the DR in this caption.}
\label{tab:observations}
\end{table*}

The channelized time series data generated by the uGMRT were recorded using the GMRT Wideband Backend \citep[GWB:][]{rkg+2017} in a binary raw data format, along with the timestamp at the start of the observation in a separate ASCII file.
We used a pipeline named \pinta{} \citep{smj+2021} to convert these raw data files into partially folded  \psrfits{} archives, which can be further analyzed using popular pulsar softwares such as \psrchive{} \citep{hsm2004}.
The time series data were folded using pulsar ephemerides derived from the IPTA DR2 \citep{pdd+2019}.
\pinta{} accommodates two different methods for RFI mitigation, out of which we use \rficlean{} \citep{mlv2020} in this work.

All profiles are pre-processed by time-collapsing and applying frequency offset corrections\footnote{See \citet{smj+2021} for discussion on the frequency labels for uGMRT data implemented in \pinta{}. For uGMRT observations carried out between 18/10/2019 and 29/07/2021 using the TGC control system, the frequency labels require an additional +10 kHz correction.} before ToA and DM estimations.
Additionally, backend delays introduced by the GWB that depend on the observation settings, described in \citet{rks+2021}, must be corrected for to achieve high-precision  timing.
These corrections are incorporated in the \psrfits{} archive headers using the \texttt{be:delay} field\footnote{These offset corrections are implemented in the latest version of \pinta{} used in this work, although it was not present in the earlier version.}. The \texttt{pat} and \texttt{pptoa} commands\footnote{See Sections \ref{sec:narrowband} and \ref{sec:wide-band} for more details.} apply the appropriate corrections to the ToAs generated from such archives.

\section{Methods for ToA and DM estimation and pulsar timing}
\label{sec:toas_dms}

In this section, we discuss the methods adopted for estimating the ToAs and DMs from our observations. These methods, as elaborated in the following sections, can be broadly classified into two categories: (1) narrowband, and (2) wideband techniques.

\subsection{Narrowband technique}
\label{sec:narrowband}

Precise ToA and DM estimations at low frequencies using the traditional narrowband technique are restricted by the following factors: (1) the estimated DM and frequency-dependent profile shape evolution are covariant with each other, (2) frequency-collapsing the data to improve the signal-to-noise ratio (S/N) can introduce smear if uncorrected DM variations are present, and (3) interstellar scatter broadening can introduce systematic biases in the estimated DMs.
To minimize the systematic errors introduced by factors (1) and (2), we adopt an iterative strategy that is described below.
We are currently developing strategies for minimizing the systematic bias introduced by scatter broadening that will be presented elsewhere.

\subsubsection{Template generation}
\label{sec:templgen}

In order to estimate DMs and ToAs from our observations, we begin by creating noise-free templates for each pulsar.
For this purpose, we identify a high-S/N observation for each pulsar from the latest uGMRT cycle (cycle 41), which lies outside the span of the present data release.
We do not add multiple epochs together since the uncorrected DM variations can introduce significant smear in the shape of the average profile, especially at low frequencies.
Instead, we  use the Band 3 and Band 5 profiles with their full available frequency resolution from the designated high-S/N epochs for template construction.

The DM values estimated using a template can exhibit a constant offset from the true DM of the pulsar depending on the fiducial DM that is used to dedisperse the template itself \citep{kmj+2021}.
Such offsets are taken care of by aligning the frequency-resolved template profiles across all the bands. 
This is initially achieved by adopting one of the methods described by \citet{kmj+2021}, which aligns the frequency-resolved profiles across both the bands by maximizing the S/N using the \texttt{pdmp} command of \psrchive{}.
After the preliminary alignment using \texttt{pdmp} DM from band 3, we prepare frequency-resolved templates in each band using an optimal wavelet smoothing algorithm \citep{dfg+2013} available in \psrchive{} via the \texttt{psrsmooth} command.

While estimating the narrowband ToAs, we are faced with a trade-off between the  ToA precision and frequency resolution due to the fact that a narrow frequency channel may not yield enough S/N for precise ToA determination.
Therefore, we are forced to partially frequency-collapse the profile, which will lead to profile smear, if the profile is not dedispersed to either a fitted fiducial DM value or accurate epoch-by-epoch DM variations.
As we do not have prior knowledge of the DM variations, we must apply an iterative method for accurately estimating the DMs and the ToAs.

To do this, we begin by dedispersing the Band 3 template and subintegrations from the Band 3 template epoch data itself to an initial, possibly inaccurate value of the \texttt{pdmp} DM. 
The Band 3 subintegrations are cross-correlated with this template and frequency-resolved Band 3 ToAs are generated.
An initial pulsar ephemeris, which is the same as the one derived from IPTA DR2 \citep{pdd+2019} and used for folding the data is used to fit the ToA residuals to a fiducial DM using \tempotwo{} \citep{ehm2006, hem2006}. This fitted DM obtained from the alignment of Band 3 residuals alone for the template epoch is used to dedisperse both Band 3 and Band 5 templates.

\begin{figure}
\centering
\includegraphics[width=\linewidth]{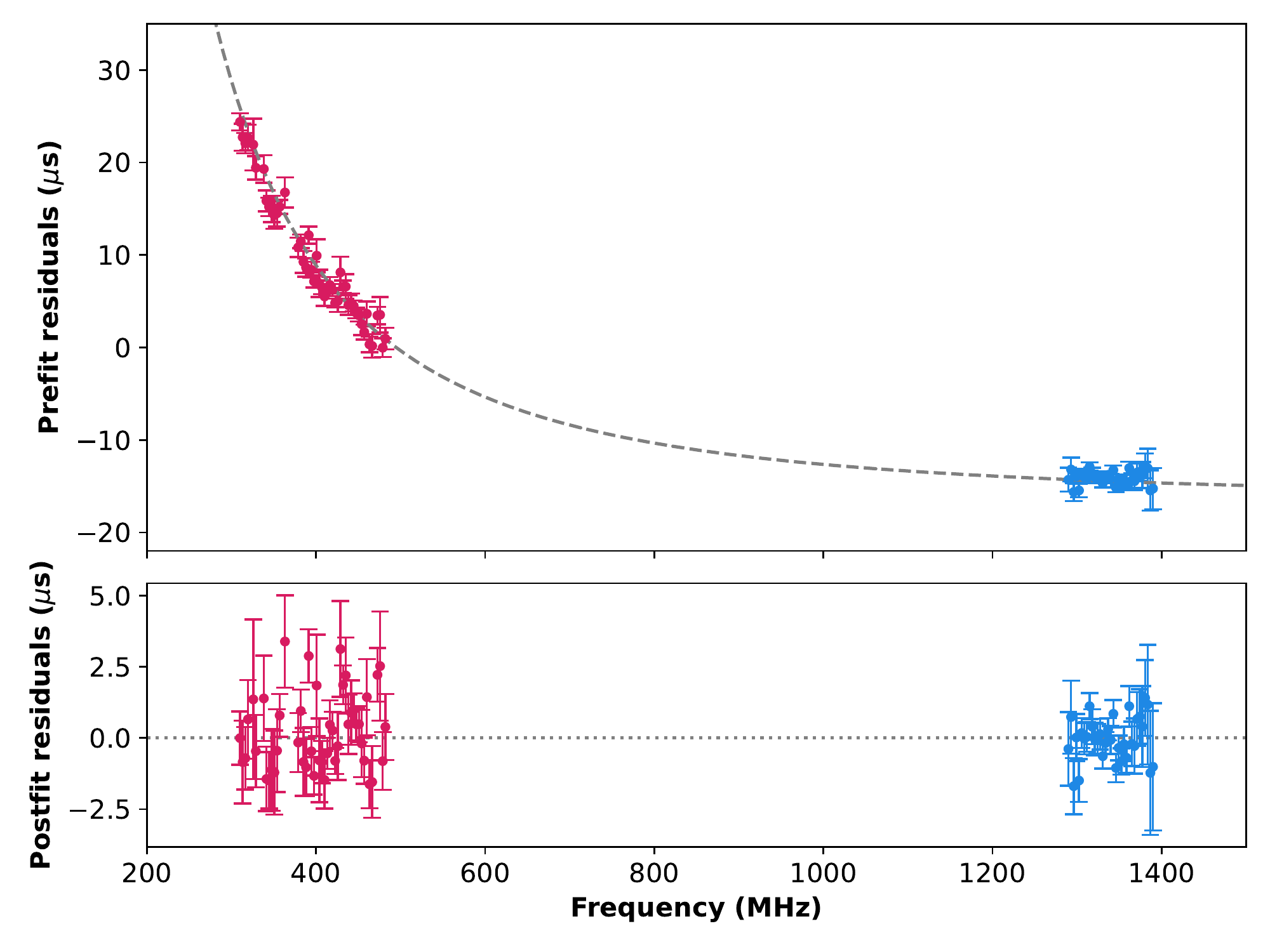} 
\caption{A sample figure illustrating that we do not require FD parameters while using frequency-resolved templates to obtain ToAs. In the top panel of the figure, the ToAs obtained from the observation of PSR J1909$-$3744 simultaneously at Band 3 and Band 5 are shown and in the bottom panel, the same ToAs are depicted after removing the residual DM trend. The blue points show Band 5 ToAs and the red ones show Band 3 ToAs respectively. The dashed line in the top panel is drawn for indicative purposes only and is not a fit.  No additional trend is visible in the residuals, and hence there is no need to fit for additional FD parameters.}
\label{align}
\end{figure}

In the next step, to take full advantage of our simultaneous multi-band observations, we use the dedispersed frequency-resolved templates to generate sub-banded ToAs in Band 3 and Band 5 by cross-correlating the corresponding subintegrations from the template epoch.
The multi-band ToA residuals are aligned across sub-bands by fitting for a DM.
This final DM estimated from the template epoch aligns all the frequency-resolved ToAs in Band 3 and Band 5, thus eliminating any frequency-dependent (FD) dispersive trend in the template epoch as demonstrated in Figure \ref{align}.
The iteratively estimated DM is applied across all the data in both the bands as the new fiducial value. The final templates are dedispersed using this fiducial DM estimate.
Since the cross-correlation between the data and template with full frequency resolution takes place across sub-bands, any frequency-dependent profile evolution across sub-bands is taken care of entirely by the frequency-resolved templates. As a result, fitting for additional FD parameters \citep{abb+2015} is not required in our analysis.

\subsubsection{ToA and DM estimation}
\label{sec:iterative}

\begin{figure*}[ht]
    \centering
    \includegraphics[width=\linewidth]{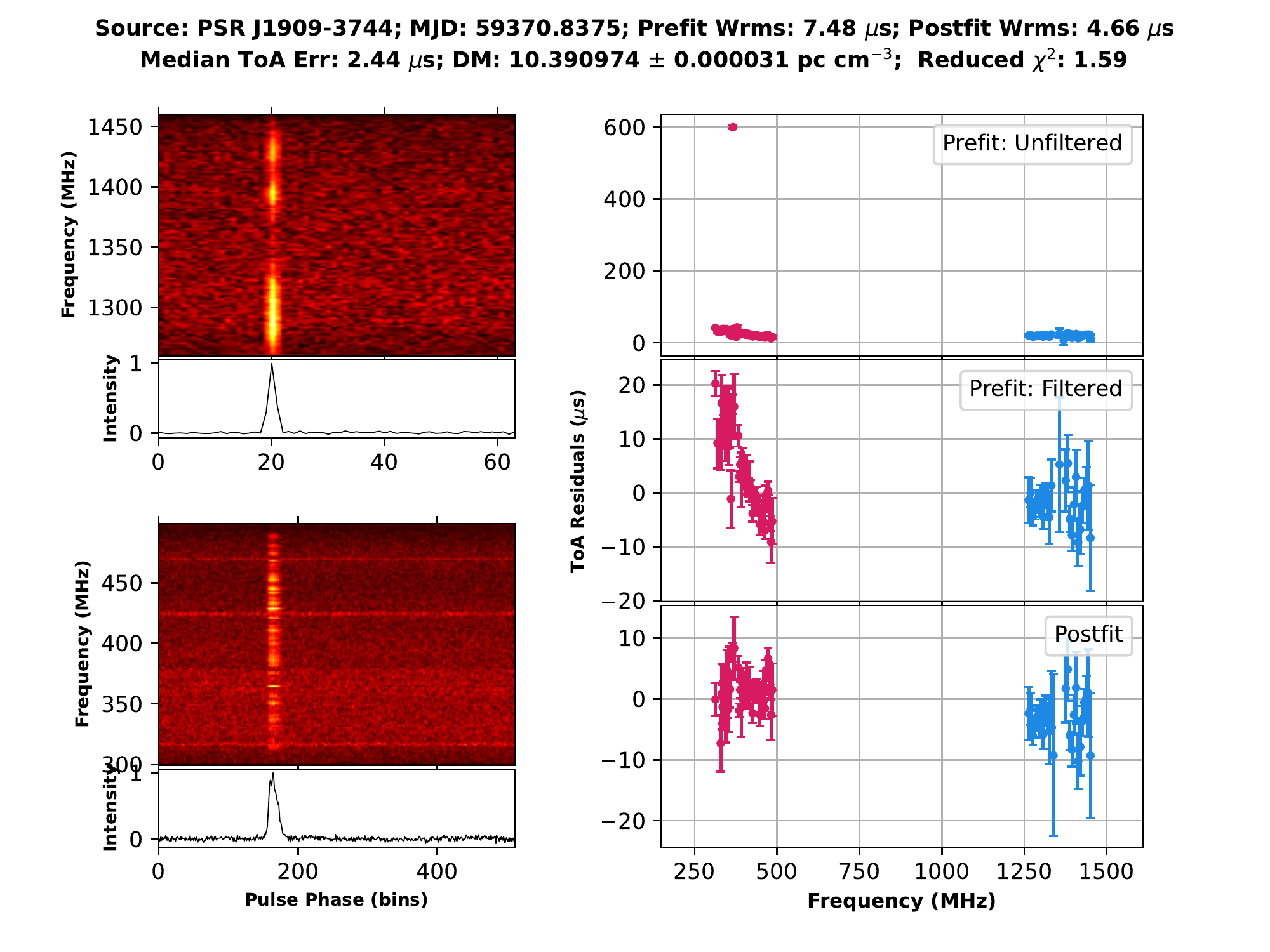}
    \caption{A sample figure for the \dmcalc{} analysis of PSR J1909$-$3744. The panels on the right side show the different steps in the process. The top panel shows the raw ToAs as obtained from \texttt{ArrivalTime} class in \texttt{PSRCHIVE}. Clearly, there is an outlier ToA present in the Band 3 ToAs, which got removed by using Huber regression analysis and the resulting ToAs are shown in the middle panel. After performing DM fit across the whole frequency range, the resultant DM corrected ToAs are shown in the bottom panel. On the Left-hand side of the plot, the profiles across the frequency range as an image as well as the time and frequency averaged profile are shown for each observing band. The details of the fit and other parameters are shown at the top of the plot.}
    \label{dmcalc_illustration}
\end{figure*}

We then run \dmcalc{} \citep{kmj+2021} on all the available band profiles together using the fitted fiducial DM and the frequency-aligned templates, to obtain the epoch-by-epoch DM variations for each pulsar.
\dmcalc{} uses the Python interface of \psrchive{} to obtain frequency resolved ToAs. It then uses the Huber regression \citep{huber1964} to fit for a quadratic trend on the ToA residuals (obtained with \tempotwo{}) and remove ToAs that are beyond 3$\sigma$ of the median absolute deviation of the residuals. After this step, the remaining ToAs are then passed on to \tempotwo{} for DM fitting. The resulting DM and frequency-resolved ToAs with IPTA-specific flags are written out into ASCII files for timing.
A sample analysis plot produced by running \dmcalc{} on the observation of PSR J1909$-$3744 on MJD 59370 is shown in Figure~\ref{dmcalc_illustration}. It can be seen that the outlier rejection methodology used in \dmcalc{} helps remove large outliers and aid in improving the DM measurement.
The initial estimate of the number of sub-bands for partial frequency-collapsing of the data is obtained using qualitative arguments to make the median ToA errors in both bands approximately equal. These numbers are refined further with preliminary timing and after carefully checking for any residual frequency evolution of the profiles across the sub-bands for each pulsar.
This refined estimation of the optimal number of sub-bands to be used in each band for each available bandwidth (100 or 200 MHz depending on the corresponding observation cycle), is used for a final run of \dmcalc{} on all the data.
The resulting ToAs are then used to obtain the final narrowband timing solutions using \tempotwo{}.

\subsection{Wideband technique}
\label{sec:wide-band}

The high precision measurement of DMs and ToAs are crucial to the PTA experiments for nanohertz gravitational wave detection. 
Various PTAs are now using wideband receivers and associated backends while implementing real-time coherent dedispersion \citep{gak+2017,hmd+2020}, to acquire more precise ToAs. Such wideband measurement helps in incorporating the effects of pulse profile evolution with frequency and DM variations. Hence, it is essential to use a method that estimates ToAs, while including a frequency-dependent model of the average pulse profile. \citet{pdr2014} and \citet{ldc+14} first provided algorithms for the simultaneous measurement of DMs and ToAs from wideband pulsar data, called the "wideband timing technique". The algorithm proposed in \citet{pdr2014} has been developed into a comprehensive package named \pulseportr{}. Further details regarding this technique can be found in \citet{p2019}. The application of this technique has been undertaken on various datasets \citep{aab+2020b,fcp+2021,nag+2022,kbd+2022,srb+22}. 
\par
Here, we apply this technique on the InPTA uGMRT dataset of 14 pulsars for estimating the DMs and ToAs.  We also compare our results with that of the narrowband technique described in Section \ref{sec:narrowband}. The uGMRT provides a unique advantage to study the effects of ISM in the low-frequency (300-500 MHz) range. It was recently shown in \cite{nag+2022} that high-precision DMs and ToAs can be obtained using wideband analysis of Band 3  uGMRT observations. Hence, we focus on carrying out the wideband analysis only in Band 3. The implemented procedure is as follows:
\begin{enumerate}
    \item \textit{Fiducial DM}: We dedisperse the data from all the epochs, including the template used for cross-correlation, using the fiducial DM estimated in the narrowband technique using the iterative procedure described in Section \ref{sec:templgen}. This is to ensure that the narrowband and wideband procedures use the same fiducial value of DM.
     \item \textit{Number of frequency channels}: As mentioned earlier, during the uGMRT Cycles 34 and 35, the InPTA observations were carried out with 100 MHz bandwidth, whereas the later cycles used 200 MHz bandwidth observations. The number of channels (sub-bands) used for the InPTA observations is usually large (see Table \ref{tab:observations}), and some of these sub-bands may contain a very weak pulsed  signal. So, we choose to partially collapse the number of sub-bands such that there is reasonable S/N in each sub-band. On the other hand, we also try to retain the information on profile evolution by not using a significantly low number of sub-bands. Hence, the optimal number of sub-bands determined in our analysis for different pulsars are 16, 32, or 64 for 200~MHz data, and 16 or 32 for 100~MHz data.
     
     \item \textit{Template generation}: We use the same epoch data file for template generation as used in the narrowband analysis. Upon careful investigation, we also found that it is important to excise frequency channels with any residual RFI from the template epoch, otherwise, we see noisy eigenprofiles upon PCA decomposition of the frequency-resolved template epoch data in the wideband analysis. Hence, we excise the RFI channels from the template epoch using \texttt{pazi} command of \texttt{PSRCHIVE} package.
     
     \item \textit{Number of eigenprofiles}: For each pulsar, we determined the number of eigenprofiles that are required in the wideband analysis to accurately model the profile evolution. The procedure we followed to select the number of eigenprofiles includes a careful check of the following four points - a) we first look at the eigenprofiles and choose an initial guess for the number of eigenprofiles including all eigenprofiles, which show significant pulsed S/N; b) we then select the most dominant normalized cumulative eigenvalues augmenting the number chosen in step (a). This ensures that the maximum amount of information on profile evolution is included with the selected eigenprofiles; c) we also make sure that the eigenvectors show significant variation with respect to the radio frequency by checking the parameterized coordinate curve vs. frequency plot [refer to Figure 4 and Equations 5 \& 7 of \citet{p2019}] for all the eigenprofiles; d) finally, we confirm that the DM estimate saturates or matches within the 1 sigma range for the selected number of eigenprofiles for 3 randomly chosen epochs. This reaffirms that the choice of the number of eigenprofiles is reasonable to account for the profile evolution.
     \item \textit{DM and ToA generation}: Once the number of eigenprofiles are determined, the ToAs and DM timeseries are generated using the \texttt{pptoas} module of \pulseportr{}.
\end{enumerate}

\subsection{Pulsar timing}
\label{sec:timing}

The ToAs obtained in Section \ref{sec:iterative} are measured using a local topocentric time standard provided by the hydrogen maser clock at the GMRT observatory.
As the GMRT observatory clock standard (UTC(gmrt)) is synchronized with the Global Positioning System clock standard (UTC(GPS)), no clock correction is needed for the UTC(gmrt) $\rightarrow$ UTC(GPS) conversion.
The ToAs were further transformed into Terrestrial Time as defined by the Bureau International des Poids et Mesures (TT(BIPM2019)), after accounting for the variable rotation of the Earth using data published by the International Earth Rotation and Reference Systems Service (IERS).
Finally, the ToAs are transformed from TT(BIPM2019) to the Barycentric Coordinate Time \citep[TCB:][]{k2005} after correcting for the orbital motion of the Earth in the solar system.
Corrections for the solar system propagation delays, to move the ToAs from a topocentric reference frame to the  Solar System Barycentre (SSB) frame, were applied based on the DE440 solar system ephemeris published by NASA JPL \citep{pfwb2021}.
A detailed discussion on the various clock corrections may be found in \citet{hem2006}. 
The final ToAs obtained were then fitted using \tempotwo{} to obtain the timing residuals after applying the clock and barycentric corrections as described above. 
We use the pulsar ephemerides derived from IPTA DR2 \citep{pdd+2019} and update the DMEPOCH and DM with the corresponding values for the template epoch obtained in the \dmcalc{} run. This updated ephemerides file is used to perform the timing analysis and obtain timing solutions. The timing procedure involves epochwise DM correction by incorporating the \dmcalc{} DM time series obtained from our low-frequency multi-band data. The application of the estimated DM time series in the pulsar ephemeris is done via DMX parameters which correspond to a piecewise-constant DM model. We use the DMXs values calculated from the \dmcalc{} run in our pulsar ephemeris file and fit for the DMXs. Additionally, we incorporated our \dmcalc{} DM time series using DMMODEL which corresponds to a linear spline DM model \citep{kcs+2012}. After comparing the two different approaches and following a rigorous consistency check, it was observed that fitting the DMX parameters for selected epochs leads to better post-fit weighted RMS residuals and reduced chi-squares. The DMX approach also ensures consistency with the wideband timing procedure which fits for DMX parameters. Thus, we used the DMX approach to incorporate the \dmcalc{} DM time series in our timing procedure. The narrowband timing involved fitting for a certain spin, binary or astrometric parameters depending on the pulsar. In most cases, we need not fit for the astrometric, binary delay \citep{dd1986} and Kopeikin delay \citep{k1995,k1996} parameters (where applicable), as our dataset is still not sensitive enough to constrain these better than the IPTA DR2 \citep{pdd+2019}. 
However, if any corresponding trends in the timing residuals were seen which could signify uncorrected pulsar parameters, then we fitted for only those respective parameters.
There could be various other sources of uncertainties that have not been taken into account in the ToA generation method discussed in \ref{sec:iterative}. Such processes could introduce additional white noise to the estimated ToA uncertainties. This additional white noise is accounted for by scaling the estimated ToA uncertainties with EFAC parameters.

The wideband timing residuals are generated using the wideband likelihood method described in Appendix B of \citet{aab+2020b}. This likelihood is implemented in the \texttt{TEMPO} \citep{nds+2015} pulsar timing software package. The wideband DM measurements from the ToAs are used as priors on the DM model parameters. The par file obtained in the first step of the previous method is transformed to a \texttt{TEMPO} compatible par file using \texttt{transform} plugin of \tempotwo{}. The same parameters that are fitted in narrowband timing are also fitted in wideband timing for each pulsar. The EFAC and DMEFAC parameters are used to scale the ToA uncertainties. These parameters are tuned iteratively until the reduced $\chi^2$ is close to 1.0.

\subsection{DM structure function analysis}
\label{sec:sfmethod}

Due to the relative motion of the pulsar and the Earth, the line of sight connecting the pulsar to the earth samples different portions of the ISM at different times.
This, along with the spatial fluctuations of the ISM electron density due to turbulence, leads to a stochastic time-dependence of the DM.
Inhomogeneities due to turbulence are usually modeled using a power-law spectrum of the form $P(q)=C q^{-\beta}$, where $q$ represents the spatial frequency \citep{r1990}.
For the Kolmogorov turbulence model, which is often assumed, the power law index $\beta=11/3$.
In this section, we investigate the spectral properties of the InPTA DM time series and its consistency with the DM time series obtained from the IPTA DR2 \citep{pdd+2019}.

The spectral index $\beta$ of a time series $\text{DM}(t)$ can be estimated by computing its structure function, which is defined as  \citep{rdb+2006,yhc+2007}
\begin{equation}
    \text{SF}(\tau)=\left\langle \left(\text{DM}(t+\tau)-\text{DM}(t)\right)^{2}\right\rangle,
\end{equation}
where $\tau$ is a time lag and the angular brackets represent averaging over the time variable $t$.
If the ISM electron density fluctuations follow a power law spectrum, then the structure function of the resulting DM time series will be a power law of the form $\text{SF}(\tau) = A \tau^\alpha$, where $\alpha=\beta-2$ \citep{ktr1994}.
Assuming stationary interstellar turbulence, for the IPTA DR2 \citep{pdd+2019} and InPTA DR1 DM time series to be deemed consistent, their structure functions should follow the same power law model.

We estimate the structure function of a given DM time series, namely $\text{DM}_i$ at times $t_i$ with measurement uncertainties $\epsilon_i$, as follows. 
We compute pair-wise DM differences $\Delta_{ij}:=|\text{DM}_i-\text{DM}_j|$ corresponding to time lags $\tau_{ij}:=|t_i-t_j|$, and partition $\tau_{ij}$ values into $B$ number of bins that are evenly spaced on a logarithmic scale.
The structure function at each $\tau$ bin can be computed by averaging the $\Delta_{ij}^2$ values in that bin.
Detailed descriptions of the algorithm for computing the structure function may be found in \citet{rdb+2006} and \citet{yhc+2007}. 
\citet{yhc+2007} also describes a bias introduced in the SF estimation by the DM measurement uncertainties, which is corrected for in our analysis.

The uncertainties $\sigma_i$ in the SF estimates described above arise due to two major sources. 
The first uncertainty component arises due to the DM measurement uncertainties and  can be estimated by propagating the DM measurement uncertainties \citep{yhc+2007}.
The second component is the variance of $\Delta_{ij}^2$ terms in each bin \citep{rdb+2006}.
Assuming these components to be independent for simplicity, we estimate the SF uncertainty in each bin by combining them in quadrature.
The uncertainty $\sigma_b$ will be dominated by the DM uncertainty component for shorter time lags and by the $\Delta_{ij}^2$ variance component for longer time lags.

\subsubsection{Structure function of the combined IPTA+InPTA DM time series}

The unique advantage of the uGMRT is its ability to record radio signals arriving simultaneously over widely separated frequency bands in low and high-frequency regimes (Band 3 and Band 5 in the present case).
Fitting for DM on the simultaneously acquired Band 3 and Band 5 ToAs on a certain epoch enables us to achieve the highest precision in DM estimation so far.
Such precise DM estimates are not attainable in the IPTA DR2 \citep{pdd+2019} as its observations were taken only at relatively high radio frequencies and are therefore less sensitive to DM measurements as compared to the uGMRT observations.
Moreover, the IPTA DR2 DM time series is obtained using the method described in \citet{kcs+2012} that models the DM as a linear spline in time.
This method implicitly smooths over epoch-wise DM variations, and the resulting time series has a cadence that is longer than the underlying pulsar observations.
Hence, the InPTA DR1 and IPTA DR2 DM time series are, in principle, complementary for estimating the structure function where the InPTA data provide the short-lag components through epoch-by-epoch DM measurements whereas the IPTA data provides the large-lag components by having a long time baseline.
Therefore, it is desirable to combine the two DM time series while estimating the DM time series spectral parameters through structure function analysis.

Unfortunately, simply combining the two time series is not enough as there is an unknown constant offset between them, arising from the different fiducial DMs used while generating the ToAs.
Therefore, we  fit for the offset along with the power law parameters, as described below. Alternatively, one could also estimate this offset by obtaining a phase-connected timing solution across the two datasets. We shall investigate this in a future work.

Let the DM offset between the two datasets be $a$. 
We estimate the combined structure function given some $a$, $\text{SF}(\tau;a)$ by (1) adding this offset to the second DM time series, (2) appending them together, and (3) computing the structure function of the resulting longer time series. 
We then fit this combined structure function for the power law model parameters $A$ and $\alpha$, together with the offset $a$, by minimizing the following weighted least squares metric:
{\small
\begin{equation}
    \chi^{2}(A,\alpha,a)=\sum_{i=1}^B\frac{\left((\log_{10} A + \alpha \log_{10}\tau)-
    \log_{10} \text{SF}(\tau_{i},a)\right)^{2}}{{\sigma'}_{i}^{2}}\,,
\end{equation}
}
where $\sigma'_{i}$ is the uncertainty in $\log_{10} \text{SF}(\tau_i,a)$ that is related to $\sigma_i$ as $\sigma'_{i}=\sigma_{i}/(\text{SF}(\tau_i;a)\ln 10)$, and the summation is over the $\tau$ bins. 
We can then minimize $\chi^{2}(A,\alpha,a)$ to get the optimal value of $a$ along with the desired fit parameters $\log_{10} A$ and $\alpha$.

\section{Results and discussion} 
\label{sec:result}

\subsection{DM time series}

\begin{table*}
\begin{tabular}{|c|c|c|c|}
\hline
\hline
\textbf{\begin{tabular}[c]{@{}c@{}}Pulsar\\Name\end{tabular}} &
\textbf{\begin{tabular}[c]{@{}c@{}}Band 3 NB (pc cm$^{-3}$)\\ {\begin{tabular}{cc} Median & Minimum \end{tabular}} \end{tabular}} &
\textbf{\begin{tabular}[c]{@{}c@{}}Band 3 WB (pc cm$^{-3}$)\\ {\begin{tabular}{cc} Median & Minimum \end{tabular}} \end{tabular}} &
\textbf{\begin{tabular}[c]{@{}c@{}}Band 3+5 NB (pc cm$^{-3}$)\\ {\begin{tabular}{cc} Median & Minimum \end{tabular}} \end{tabular}} \\ \hline
J0437$-$4715 & {\begin{tabular}{cc} $6.6 \times 10^{-5}$ & $8.0 \times 10^{-6}$ \end{tabular}} & {\begin{tabular}{cc} $9.8 \times 10^{-6}$ & $2.6 \times 10^{-6}$ \end{tabular}} & {\begin{tabular}{cc} $3.1 \times 10^{-5}$ & $5.0 \times 10^{-6}$ \end{tabular}} \\
J0613$-$0200 & {\begin{tabular}{cc} $1.1 \times 10^{-4}$ & $1.5 \times 10^{-5}$ \end{tabular}} & {\begin{tabular}{cc} $4.2 \times 10^{-5}$ & $2.6 \times 10^{-5}$ \end{tabular}} & {\begin{tabular}{cc} $9.0 \times 10^{-5}$ & $1.1 \times 10^{-5}$ \end{tabular}} \\
J0751$+$1807 & {\begin{tabular}{cc} $6.3 \times 10^{-4}$ & $5.6 \times 10^{-5}$ \end{tabular}} & {\begin{tabular}{cc} $2.2 \times 10^{-4}$ & $7.8 \times 10^{-5}$ \end{tabular}} & {\begin{tabular}{cc} $5.2 \times 10^{-4}$ & $4.8 \times 10^{-5}$ \end{tabular}} \\
J1012$+$5307 & {\begin{tabular}{cc} $9.5 \times 10^{-5}$ & $2.0 \times 10^{-5}$ \end{tabular}} & {\begin{tabular}{cc} $5.4 \times 10^{-5}$ & $8.9 \times 10^{-6}$ \end{tabular}} & {\begin{tabular}{cc} $7.7 \times 10^{-5}$ & $2.1 \times 10^{-5}$ \end{tabular}} \\
J1022$+$1001 & {\begin{tabular}{cc} $2.4 \times 10^{-4}$ & $1.4 \times 10^{-5}$ \end{tabular}} & {\begin{tabular}{cc} $9.9 \times 10^{-5}$ & $3.6 \times 10^{-5}$ \end{tabular}} & {\begin{tabular}{cc} $2.0 \times 10^{-4}$ & $1.2 \times 10^{-5}$ \end{tabular}} \\
J1600$-$3053 & {\begin{tabular}{cc} $3.6 \times 10^{-4}$ & $2.2 \times 10^{-5}$ \end{tabular}} & {\begin{tabular}{cc} $1.8 \times 10^{-4}$ & $4.5 \times 10^{-5}$ \end{tabular}} & {\begin{tabular}{cc} $2.5 \times 10^{-4}$ & $2.1 \times 10^{-5}$ \end{tabular}} \\
J1643$-$1224 & {\begin{tabular}{cc} $2.2 \times 10^{-4}$ & $1.5 \times 10^{-5}$ \end{tabular}} & {\begin{tabular}{cc} $1.2 \times 10^{-4}$ & $6.9 \times 10^{-5}$ \end{tabular}} & {\begin{tabular}{cc} $3.4 \times 10^{-4}$ & $1.3 \times 10^{-5}$ \end{tabular}} \\
J1713$+$0747 & {\begin{tabular}{cc} $1.2 \times 10^{-4}$ & $3.1 \times 10^{-5}$ \end{tabular}} & {\begin{tabular}{cc} $9.2 \times 10^{-5}$ & $3.2 \times 10^{-5}$ \end{tabular}} & {\begin{tabular}{cc} $6.7 \times 10^{-5}$ & $2.1 \times 10^{-5}$ \end{tabular}} \\
J1744$-$1134 & {\begin{tabular}{cc} $4.0 \times 10^{-5}$ & $7.0 \times 10^{-6}$ \end{tabular}} & {\begin{tabular}{cc} $2.5 \times 10^{-5}$ & $1.0 \times 10^{-5}$ \end{tabular}} & {\begin{tabular}{cc} $3.7 \times 10^{-5}$ & $6.0 \times 10^{-6}$ \end{tabular}} \\
J1857$+$0943 & {\begin{tabular}{cc} $4.4 \times 10^{-4}$ & $1.4 \times 10^{-5}$ \end{tabular}} & {\begin{tabular}{cc} $1.5 \times 10^{-4}$ & $8.9 \times 10^{-5}$ \end{tabular}} & {\begin{tabular}{cc} $3.5 \times 10^{-4}$ & $1.3 \times 10^{-5}$ \end{tabular}} \\
J1909$-$3744 & {\begin{tabular}{cc} $3.2 \times 10^{-5}$ & $4.0 \times 10^{-6}$ \end{tabular}} & {\begin{tabular}{cc} $1.6 \times 10^{-5}$ & $6.9 \times 10^{-6}$ \end{tabular}} & {\begin{tabular}{cc} $2.2 \times 10^{-5}$ & $4.0 \times 10^{-6}$ \end{tabular}} \\
J1939$+$2134 & {\begin{tabular}{cc} $3.2 \times 10^{-5}$ & $1.0 \times 10^{-5}$ \end{tabular}} & {\begin{tabular}{cc} $3.3 \times 10^{-6}$ & $1.2 \times 10^{-6}$ \end{tabular}} & {\begin{tabular}{cc} $3.4 \times 10^{-5}$ & $3.0 \times 10^{-6}$ \end{tabular}} \\
J2124$-$3358 & {\begin{tabular}{cc} $2.3 \times 10^{-4}$ & $1.2 \times 10^{-5}$ \end{tabular}} & {\begin{tabular}{cc} $1.2 \times 10^{-4}$ & $2.0 \times 10^{-5}$ \end{tabular}} & {\begin{tabular}{cc} $2.5 \times 10^{-4}$ & $1.2 \times 10^{-5}$ \end{tabular}} \\
J2145$-$0750 & {\begin{tabular}{cc} $1.2 \times 10^{-4}$ & $6.0 \times 10^{-6}$ \end{tabular}} & {\begin{tabular}{cc} $3.3 \times 10^{-4}$ & $8.4 \times 10^{-6}$ \end{tabular}} & {\begin{tabular}{cc} $1.0 \times 10^{-4}$ & $6.0 \times 10^{-6}$ \end{tabular}} \\ \hline
\end{tabular}
\caption{Table of uncertainties in estimated DMs. The first column specifies the pulsars. The second column lists the median and minimum errors in the DM estimation using narrowband (NB) residuals obtained from ToAs in Band 3 with 200 MHz bandwidth. The third column represents similar median and minimum errors for DMs estimated using wideband (WB) technique on Band 3 data having bandwidth of 200 MHz. The fourth column enlists median amd minimum DM errors for estimations from Band 3 and Band 5 narrowband (NB) ToAs combined with bandwidth of 200 MHz in each band.}
\label{tab:results1}
\end{table*}

\begin{figure*}[h!]
    \centering
    \includegraphics[scale=1.1, width=\linewidth]{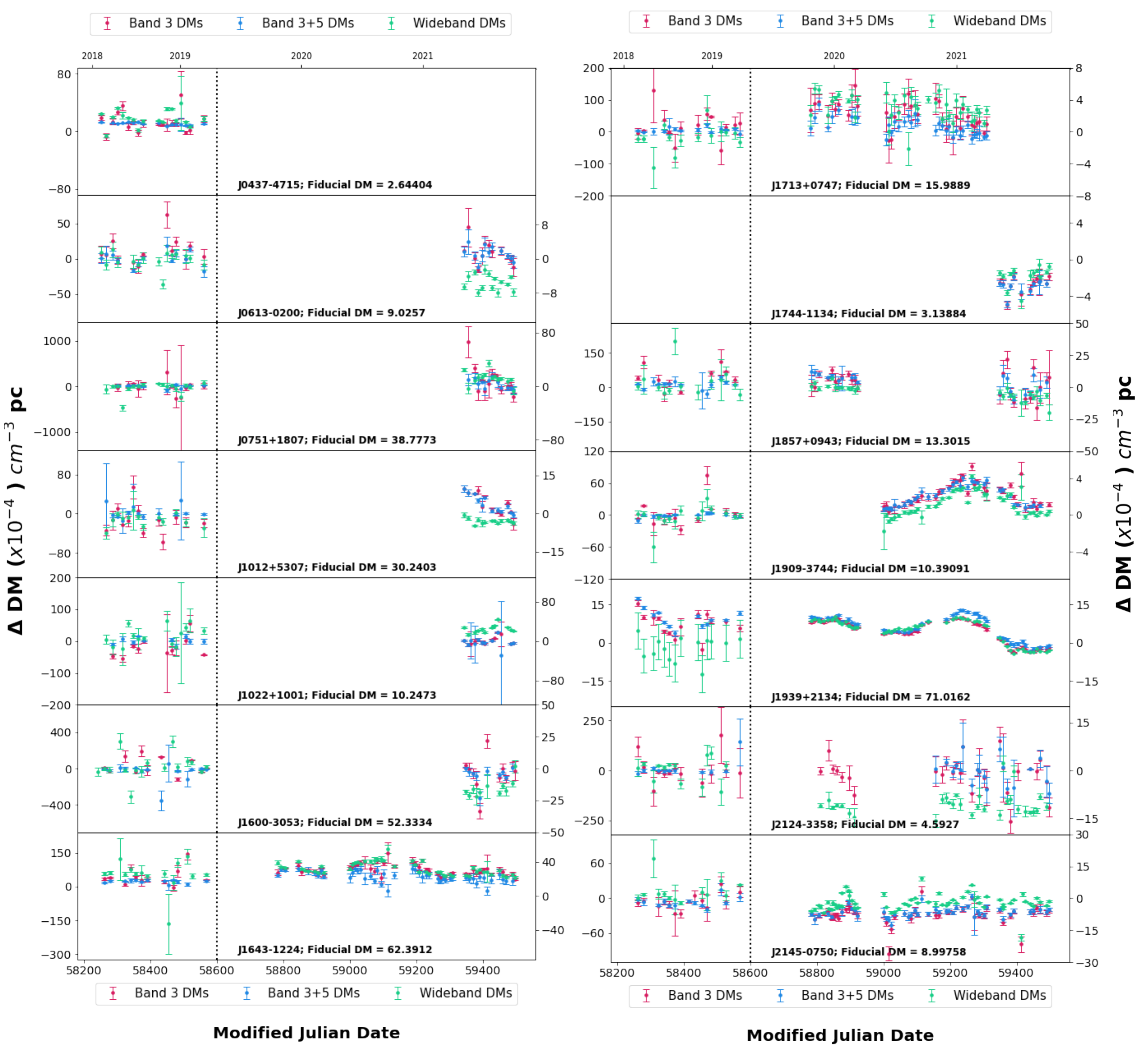}
    \caption{DM time-series for 14 pulsars. The differences ($\Delta$DM in units of $10^{-4}$ cm$^{-3}$ pc) between the fiducial DM and the corresponding estimated DMs for each pulsar estimated by (i) fitting only Band 3 narrowband ToAs (red points), (ii) fitting Band 3 and Band 5 narrowband ToAs together (blue points) and (iii) by applying the wideband technique on Band 3 data (green points), are overlaid for comparison. The values depicted on the vertical axis are DMs relative to the fiducial DMs for the respective pulsars that are obtained using the iterative method described in Section \ref{sec:templgen}. Since the precision of DM estimation from 200 MHz bandwidth data is higher than that from 100 MHz bandwidth data, the horizontal axes are split into two parts at MJD 58600 with dotted vertical lines, where epochs on the left side of the dotted line in each panel represent 100 MHz bandwidth, and epochs on the right side of the dotted vertical lines in each panel represent 200 MHz bandwidth. The vertical axes in each panel are also scaled differently for 100 MHz bandwidth (left axis) and 200 MHz bandwidth (right axis) epochs such that the DM variations are clearly visible. Pulsar names and their respective fiducial DM values are mentioned at the bottom of each respective panel. The time span in terms of the years is also denoted at the top for convenience.}
    \label{fig:dm_combined}
\end{figure*}

\begin{figure*}[h!]
\centering
\includegraphics[scale=1.0, width=\linewidth, trim={0.5cm 0 0.5cm 0},clip]{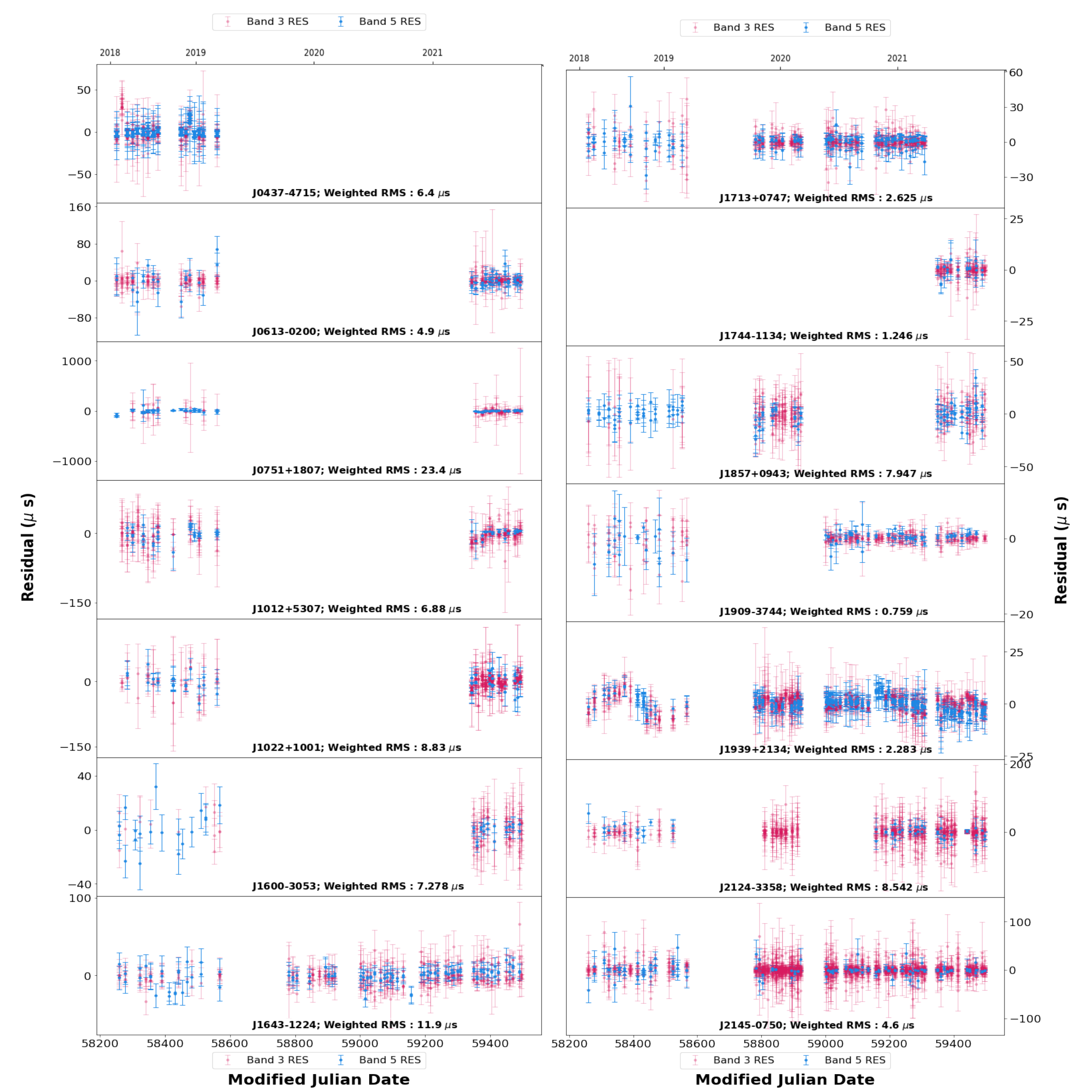}
\caption{Narrowband timing residuals for 14 pulsars. The timing residuals obtained from Band 3 and Band 5 data using the narrowband timing technique for 14 pulsars are plotted against corresponding epochs. Red points represent Band 3 and blue points represent Band 5 residuals. Pulsar names and their respective post-fit weighted RMS residuals are mentioned at the bottom of the respective panels. Epochs in terms of Modified Julian Date are depicted on the consolidated horizontal axes at the bottom. The corresponding years are also shown on the horizontal axes at the top of the consolidated panels for convenience.}
\label{residuals_comb}
\end{figure*}

In this section, we present the DM time series and timing residuals for the 14 pulsars. Table \ref{tab:results1} lists the median and minimum uncertainties in DM estimated using: (i) narrowband technique upon Band 3 data with 200 MHz bandwidth, (ii) narrowband technique upon simultaneously recorded Band 3 and Band 5 data with 200 MHz bandwidth in each band, and (iii) wideband technique upon Band 3 data with 200 MHz bandwidth. The listed uncertainty values establish that these are the highest precision DM estimations made so far. It is evident that the narrowband ToAs obtained from simultaneously recorded uGMRT Band 3 and Band 5 data lead to high precision in the DM estimation. Even higher precision is obtained in general when Band 3 DMs are estimated using the wideband technique. The extension of the wideband method for widely separated frequency bands when combined together is a work in progress. Application of such a technique on the simultaneously recorded Band 3 and Band 5 data using the uGMRT is expected to make DM estimations more precise in the near future. 

A consolidated plot illustrating the epoch-by-epoch DM variations for all 14 pulsars is depicted in Figure \ref{fig:dm_combined}.
In this figure, the vertical axes represent the differences between the epoch DMs and the fiducial DMs obtained iteratively for the respective pulsars as explained in Section \ref{sec:iterative}, and also obtained using the wideband technique as explained in Section \ref{sec:wide-band}. The fiducial value for each pulsar is mentioned inside its respective panel in the figure. Since the DMs estimated from 200 MHz bandwidth data (uGMRT observation cycles 37-40) have higher precision compared to those obtained from 100 MHz bandwidth data (uGMRT observation cycles 34-35), the 100 and 200 MHz bandwidth epochs have been separated along the horizontal axis with a vertical dotted line at MJD 58600. The scaling of the vertical axes for the epochs prior to MJD 58600 is different as compared to scaling of the vertical axis for epochs after MJD 58600. This has been done to present a clearer view of the DM variations over both 100 MHz and 200 MHz bandwidth epochs.

We observe that the known phenomenon of offsets between DM estimations from narrowband and wideband techniques is caused due to the differences in smoothing of the templates in the two methods.
We have confirmed this by performing confirmatory tests on the InPTA data for PSR J1909-3744 which exhibits the minimum amount of profile evolution and scatter broadening with a distinctive sharp profile among all the pulsars in our sample.
Indications of differences in template smoothing being the reason for such offsets appear while using narrowband templates (smoothed with some optimal wavelet smoothing algorithm in \texttt{psrsmooth}) for PCA decomposition and subsequent reconstruction for wideband DM and ToA estimations.
This method leads to wideband DMs that do not show any offset with respect to the corresponding narrowband DMs.
The same results are obtained when wideband templates (RFI removed data from template epoch reconstructed after PCA decomposition by fitting splines through significant eigenprofiles) are used for DM estimation with \dmcalc{} using the narrowband technique.
The DM offset calculated from direct cross-correlation of the narrowband and wideband templates is concurrent with the observed offset between the narrowband and wideband DMs estimated from the InPTA data for this pulsar.

PSR J0437$-$4715 was observed only in cycles 34 and 35 with 100 MHz bandwidths in Band 3 and Band 5. We have restarted monitoring the pulsar only very recently in both bands with 200 MHz bandwidth. The 200 MHz bandwidth data for this pulsar belongs to the latest observational cycle of the uGMRT and is not a part of the present data release. Similarly, PSR J1744$-$1134 was added to the InPTA sample only in uGMRT observation cycle 40. Outliers are checked for artefacts and are subject to further investigations. PSR J1713$+$0747 underwent an abrupt profile change between MJDs 59320 and 59321 \citep{ssj+21, lam21}. The post-event data for PSR J1713$+$0747 has been excluded from this release and is subject to further analysis. Interesting offsets between Band 3 and Band 3+5 DM estimations are observed in PSR J1939$+$2134. These offsets could be indicative of variable scattering in this pulsar. However, a confirmation of such a hypothesis demands an extensive study into the effects of variable scattering on simulated pulsar data. It is a question being addressed in some of our ongoing projects and is beyond the scope of the present work.

\subsection{The timing residuals}

A similar consolidated plot of the narrowband timing residuals from both bands for all 14 pulsars is shown in Figure \ref{residuals_comb}.
We also performed rigorous consistency checks in order to compare the timing solutions obtained from narrowband and wideband timing techniques. Although the two methods are different in nature, it is expected that they should give similar timing solutions. \texttt{TEMPO} (used for wideband timing) and \texttt{TEMPO2} (used for narrowband timing) are exclusively different in the supported format of the pulsar ephemeris files. Therefore, in order to compare the narrowband and wideband timing solutions, the final par files obtained from the two techniques were made commensurable using the \texttt{transform} plugin of \tempotwo{}. The fitted parameters were then compared by taking a difference, which was normalised using the quadrature sum of the uncertainties in the corresponding parameters estimated using each technique.

\begin{figure}[h!]
\centering
\includegraphics[width=\linewidth]{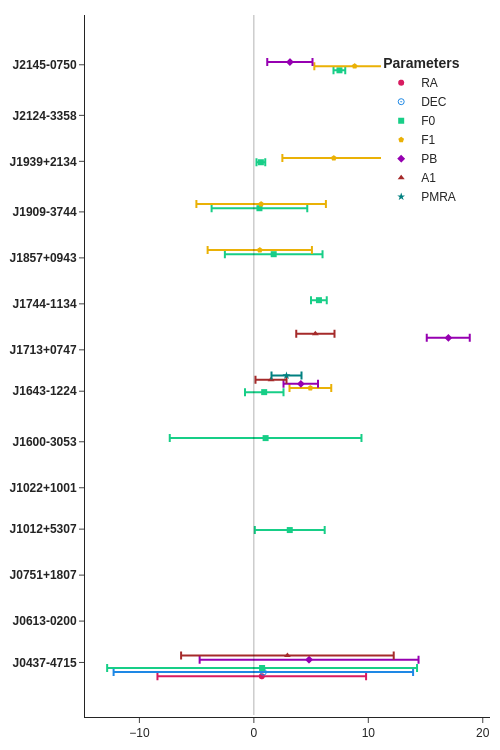} 
\caption{Comparison of fitted pulsar parameters obtained with narrowband and wideband timing. The parameter differences have been normalised using the quadratic sum of the two errors ($\sigma_{NB}$ : the error from narrowband timing, and $\sigma_{WB}$ : the error from wideband timing). The various points in the plot represent these differences whereas lengths of the error bars in the plot are given by $\sigma_{WB} / \sigma_{NB}$. For most of the pulsars, only the spin parameters have been fitted. For J0613$-$0200, J0751+1007, J1022+1001 and J2124$-$3358, none of the pulsar parameters required fitting.}
\label{consistency}
\end{figure}

A comparison between the narrowband and wideband timing solutions is presented in Figure \ref{consistency}. The plot lists the respective astrometric, spin, and binary parameters that required fitting for a given pulsar. The points on the plot depict the differences between the fitted values obtained from narrowband and wideband methods normalised in units of the quadratic sum of the uncertainties in the two methods. The horizontal error bars across the points signify the ratio of the errors in the  estimation using the wideband technique to the narrowband technique.  It is evident that for most pulsars, only the spin parameters required fitting and that the narrowband and wideband timing solutions are generally consistent with each other.

Detailed plots of the DM time series and timing residuals (estimated using both narrowband and wideband methods) for each pulsar are included in \ref{sec:appendix}.

\subsection{DM structure function analysis: Results} 
\label{sec:dmstructfunc}

\begin{figure*}[h!]
\includegraphics[width=\linewidth]{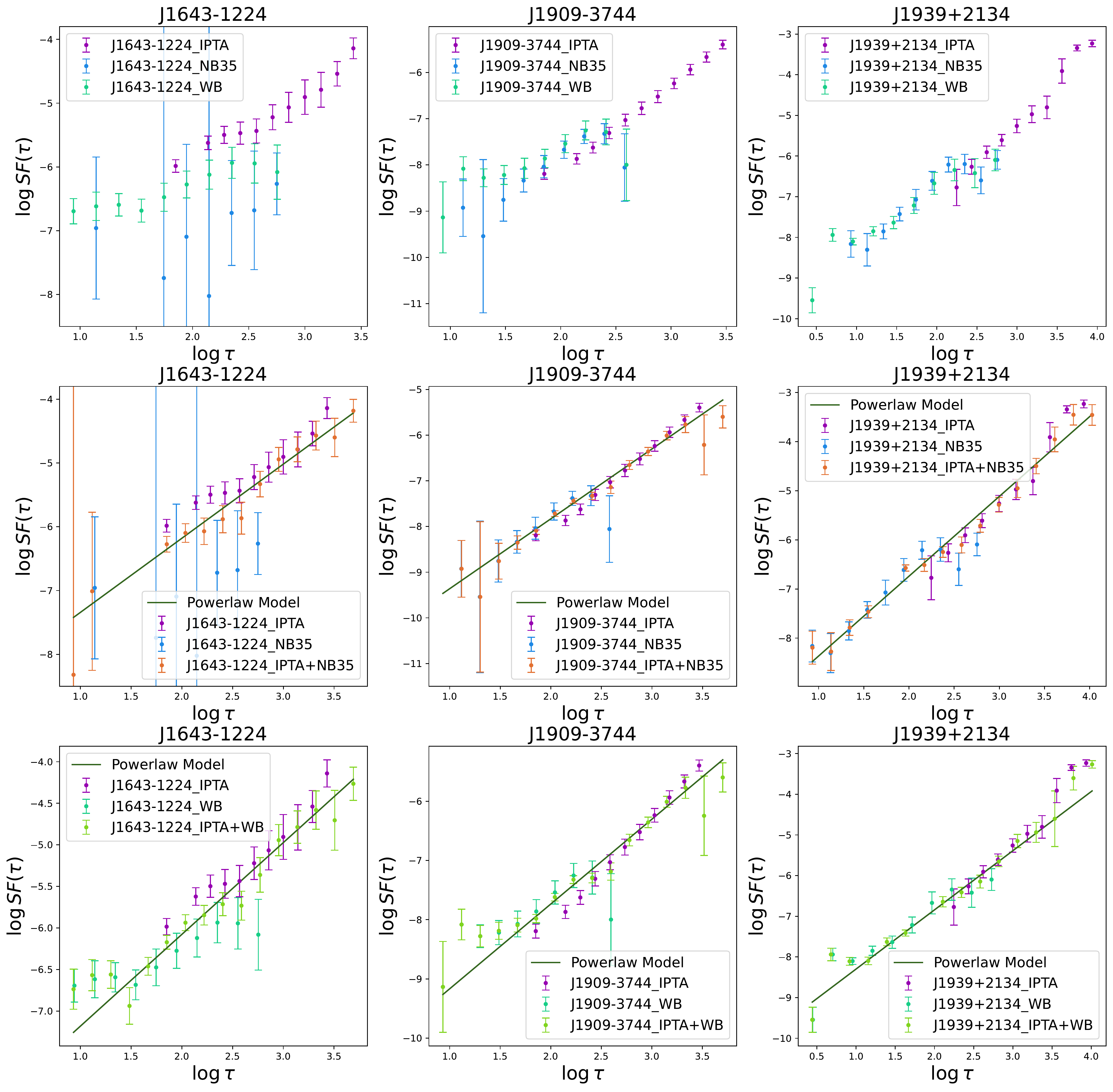}
\caption{Structure function (SF) plots for three pulsars PSR J1643$-$1224, PSR J1909$-$3744, PSR J1939$+$2134. The plots in the first row show SFs of IPTA DR2 DMs, InPTA narrowband DMs (estimated from Band 3 and Band 5 ToAs combined), and InPTA wideband DMs (Band 3). Plots in the second row show SFs obtained separately from IPTA DR2 DMs and InPTA narrowband DMs (from Band 3+5 ToAs combined), and SFs obtained upon combining the IPTA DR2 and InPTA narrowband DMs. Plots in the third row show SFs obtained separately from IPTA DR2 DMs and InPTA wideband DMs (Band 3), and SFs obtained from the combination of IPTA DR2 and InPTA wideband DMs. Note that the InPTA DMs used (both narrowband Band 3+5 and wideband Band 3) belong to epochs observed with a bandwidth of 200 MHz in each band.}
\label{fig:sfplot}
\end{figure*}

In this section, we present the structure functions obtained for three InPTA pulsars, namely PSRs J1643$-$1224, J1909$-$3744, and J1939$+$2134, following the procedure outlined in Section \ref{sec:sfmethod}.
These three pulsars are known to exhibit strong DM variations, allowing us to estimate their structure functions despite the relatively short time span of our data.
Due to this short time span, we were unable to obtain meaningful results for the remaining pulsars.

Typically, a structure function may contain three distinct regions: (1) a noise region at smaller time lags dominated by measurement noise, (2) a saturation region at larger time lags due to a finite sampling of the DM time series, and (3) a structure region in the middle that typically follows a power-law model.
We exclude the noise and saturation regions from our analysis where they are identified since they can bias the estimation of the model parameters.

We present in Figure \ref{fig:sfplot} the structure functions for the three pulsars (three columns) estimated from the InPTA narrowband (Band 3+5) and the wideband (Band 3) DM time series presented in the previous subsection.
We exclude the DM estimates from the observation cycles 34 and 35 as they are of much lower precision due to the shorter observation durations and a lower bandwidth.
The first row of Figure \ref{fig:sfplot} shows a comparison between the InPTA narrowband and wideband structure functions alongside their IPTA DR2 \citep{pdd+2019} counterparts.
The second and third rows show the structure functions obtained by the combination of the InPTA narrowband and wideband DM time series respectively with the IPTA DM time series together with the corresponding best-fit power-law models.
The best-fit parameters ($\log_{10} A$, $\alpha$, and $a$) are listed in Table \ref{tab:sf-bestfit}.

\begin{table*}[h!]
\caption{Structure function best-fit parameters}
\label{tab:sf-bestfit}
\begin{tabular}{|c|ccc|ccc|}
\hline\hline
\textbf{Pulsar} & \multicolumn{3}{c|}{\textbf{NB+IPTA}}                                               & \multicolumn{3}{c|}{\textbf{WB+IPTA}}                                               \\ %\cline{2-7} 
  \textbf{Name}                   & {$\boldsymbol{\log_{10} A}$}  & {$\boldsymbol\alpha$} & $\boldsymbol{a}$        & {$\boldsymbol{\log_{10} A}$}  & {$\boldsymbol\alpha$} & $\boldsymbol{a}$        \\ \hline
J1643$-$1224           & \multicolumn{1}{c}{-8.5(5)} & \multicolumn{1}{c}{1.16(8)}  & 0.014(5)   & \multicolumn{1}{c}{-8.3(2)} & \multicolumn{1}{c}{1.01(8)}  & 0.01(1)   \\ %\hline
J1909$-$3744           & \multicolumn{1}{c}{-10.9(2)} & \multicolumn{1}{c}{1.53(6)}  & 0.000(3) & \multicolumn{1}{c}{-10.6(2)} & \multicolumn{1}{c}{1.43(8)}  & 0.000(3) \\ %\hline
J1939$+$2134           & \multicolumn{1}{c}{-10.0(3)} & \multicolumn{1}{c}{1.6(2)}  & 0.111(3)   & \multicolumn{1}{c}{-9.75(9)} & \multicolumn{1}{c}{1.45(4)}  & 0.104(8)   \\ \hline
\end{tabular}
\end{table*}

From Figure \ref{fig:sfplot}, we can clearly see that the structure functions estimated from the InPTA DR1 narrowband and wideband DM time series are consistent with each other within $1\sigma$ error bars for the three pulsars.
The InPTA structure functions show saturation regions for PSRs J1909$-$3744 and J1939$+$2134 and possible noise regions for PSRs J1643$-$1224 and J1909$-$3744.
Excluding the saturation regions, the InPTA structure functions (both narrowband and wideband) are consistent with the IPTA DR2 structure functions within 2-sigma error bars where they overlap in the $\tau$ axis.
This observation may be interpreted as showing the consistency between the InPTA and IPTA DM time series, despite there being no overlap in time between the two datasets and the differing methodologies using which the DM time series were obtained (namely, epoch-by-epoch DMs vs linear spline model).
In addition, Figure \ref{fig:sfplot} clearly shows that the epoch-wise DM estimates from simultaneous multi-band observations provided by the InPTA observations, even with their short time-lag coverage, are complementary to other PTA efforts for modeling DM noise despite its shorter time baseline.

Given the short span of our data and the a priori unknown offset between the InPTA and IPTA DM time series, rigorous investigations in this direction will become possible as more data gets collected and the InPTA data is combined with the other PTA datasets for the third IPTA data release.
In addition, it will be illuminating to investigate the behaviour of the DM structure functions at shorter lags ($\sim$1 day) by performing daily cadence uGMRT observations of the above three pulsars over a 10-15 day time span.

\subsection{Investigation of frequency dependence of DM}
\label{sec:fd}

The small-scale electron density variations in the ISM will scatter the rays resulting in multi-path propagation. It was suggested by \citet{css2016} that this will affect the ToAs at higher frequencies while combining them with lower frequency ToAs as this chromatic effect depends strongly on the observing frequency. Using low-frequency observations, it was shown by \citet{dvt+2019} that the DMs are different even within about a bandwidth of 70~MHz at low frequencies and vary over several years. In a recent study by \citet{kbd+2022}, a possible PTA candidate PSR J2241$-$5236 was shown to have DM chromaticity over a wide frequency range. Since our data also spans a wide frequency range, we have performed a preliminary analysis for checking the chromaticity of DM in our data set.

As we have two widely separated bands (with wide bandwidths), we used both narrowband and wideband techniques for estimating the DMs using different combinations. 
We chose a set of pulsars (J0437$-$4715, J1643$-$1224, J1713$+$0747, J1939$+$2134, and J2145$-$0750) for which high S/N observations are available with 200~MHz bandwidths. 
Then, using wideband technique, we estimated the DM from Band 3 and Band 5 after splitting each of the bands into two 100~MHz sub-bands as well as using full Band 5 with two halves of Band 3 alone. In addition, using narrowband method, we tried four different combinations of 100~MHz portions of the data at both bands to obtain DMs. These different DM time series were compared with the DM time series shown in Figure \ref{fig:dm_combined}. We did not find any significant difference in the DM time series (both temporal and frequency-wise) between these various combinations indicating that these pulsars may not show DM chromaticity and we can use the DMs and ToAs from Band 3 for high precision timing. Even though this is true for the current analysis, we will be working on refining our methods in the future and try to see if any of the pulsars we observe show frequency dependence of DM.

\section{Conclusions}
\label{sec:conclusions}

In this work, we have presented the first data release of InPTA, which includes the pulse arrival times and  DMs for 14 millisecond pulsars observed simultaneously at 300$-$500 MHz and 1260$-$1460 MHz bands  using the uGMRT. The ToAs and DMs have been calculated using two independent methods: narrowband method as well as the wideband technique. These ToAs and DMs were further used to obtain the timing solutions of these pulsars by fitting the spin, astrometric, and binary parameters in both the narrowband and wideband framework. A comparison showing a broad agreement in the results from the two different techniques has also been demonstrated. For both these methods, we also present the DM structure function and present a few preliminary tests conducted in order to investigate the frequency dependence of DM for our pulsar sample. 

The main contribution of this release is high precision DM estimates.  It is possible to minimise the DM noise from 1260$-$1460 MHz pulse arrival times by correcting these by using the DM estimates from our simultaneous observations. The pulse arrival times in this data release may also be useful in deriving better Gaussian process DM noise models as employed by EPTA and PPTA. Lastly, the direct use of DM time series in DM noise models is a work in progress. The current data sets from the MeerKat and the FAST are  also likely to be DM noise limited like other PTAs, and may benefit from this data release. 
These data are likely to improve the quality of the proposed IPTA Data Release 3 thereby helping in the detection of GWs in the near future.

Our data for 14 pulsars are now publicly available. Further monitoring of all these pulsars is in progress as part of the InPTA efforts. We have recently included 3 additional millisecond pulsars to the InPTA sample and are observing a total of 17 pulsars during the ongoing uGMRT cycle. We plan to extend the sample further with simultaneous multi-target observations using multiple phased subarrays. The Target-of-Opportunity proposals are active to follow up on transients such as glitches, profile mode changes, or extreme scattering events in any of the IPTA pulsars observable by the uGMRT. Pulsar-specific proposals with scientific goals that require a higher cadence are being discussed. The 3.5-year dataset for 14 millisecond pulsars that we make available through this first InPTA data release will be pooled along with data from other IPTA telescopes and used to search for nanohertz gravitational waves as well as a whole plethora of auxiliary science.

\section{Data availability}
\label{sec:data}

The InPTA DR1 described in this paper, consisting of ToA measurements, pulsar ephemerides, and DM measurements using narrowband and wideband techniques, will be publicly available at \url{https://github.com/inpta/InPTA.DR1} after peer review.
The IPTA DR2 is publicly available at \url{https://gitlab.com/IPTA/DR2}.

\section{Software}
\href{http://dspsr.sourceforge.net/}{\dspsr{}} \citep{sb2011}, 
\href{http://psrchive.sourceforge.net/}{\psrchive{}} \citep{hsm2004}, 
\href{https://github.com/ymaan4/RFIClean}{\rficlean{}} \citep{mlv2020},
\href{https://github.com/abhisrkckl/pinta}{\pinta{}} \citep{smj+2021},
\href{https://github.com/nanograv/tempo}{\tempo{}} \citep{nds+2015},
\href{https://bitbucket.org/psrsoft/tempo2/src/master/}{\tempotwo{}} \citep{hem2006,ehm2006}, 
\href{https://github.com/kkma89/dmcalc}{\dmcalc{}} \citep{kmj+2021},
\href{https://github.com/pennucci/PulsePortraiture}{\pulseportr{}} \citep{p2015,p2019}, 
\href{https://github.com/lmfit}{\texttt{lmfit}} \citep{nsai2014},
\href{https://github.com/matplotlib/matplotlib}{\texttt{matplotlib}} \citep{h2007},
\href{https://astropy.org}{\texttt{astropy}} \citep{astropy+2018}

\section*{Acknowledgements}
InPTA acknowledges the support of the GMRT staff in resolving technical difficulties and providing technical solutions for high-precision work. We acknowledge the GMRT telescope operators for the observations. The GMRT is run by the National Centre for Radio Astrophysics of the Tata Institute of Fundamental Research, India.
We acknowledge the use of IPTA DR2 \citep{pdd+2019} for this paper.
AS, BCJ, YG, YM, AG, LD, SuD, and PR acknowledge the support of the Department of Atomic Energy, Government of India, under Project Identification \# RTI 4002. BCJ acknowledges the support of the Department of Atomic Energy, Government of India, under project No. 12-R\&D-TFR-5.02-0700. 
%AS is supported in part by the National Natural Science Foundation of China grant No. 11988101. 
AS is supported by the NANOGrav NSF Physics Frontiers Center (awards \#1430284 and 2020265).
SH is supported by JSPS KAKENHI Grant Number 20J20509. AmS is supported by CSIR fellowship Grant number 09/1001(12656) /2021-EMR-I and DST-ICPS (T-641). KT is partially supported by JSPS KAKENHI Grant Numbers 20H00180, 21H01130 and 21H04467 and the ISM Cooperative Research Program (2021-ISMCRP-2017). YG acknowledges the support of the Department of Atomic Energy, Government of India, under project No. 12-R\&D-TFR-5.02-0700. MPS acknowledges funding from the European Research Council (ERC) under the European Union's Horizon 2020 research and innovation programme (grant agreement No. 694745). We are grateful to the anonymous referee for a thorough and comprehensive review of our manuscript.
 
\bibliography{inpta-dr1}

\onecolumn
\newpage
\appendix
\section{DM time-series and timing residuals}
\label{sec:appendix}
The DM time series and the timing residuals obtained using the narrowband and the wideband techniques on each of the 14 pulsars are presented here.

\begin{figure*}[h!]
\includegraphics[width=0.8\linewidth]{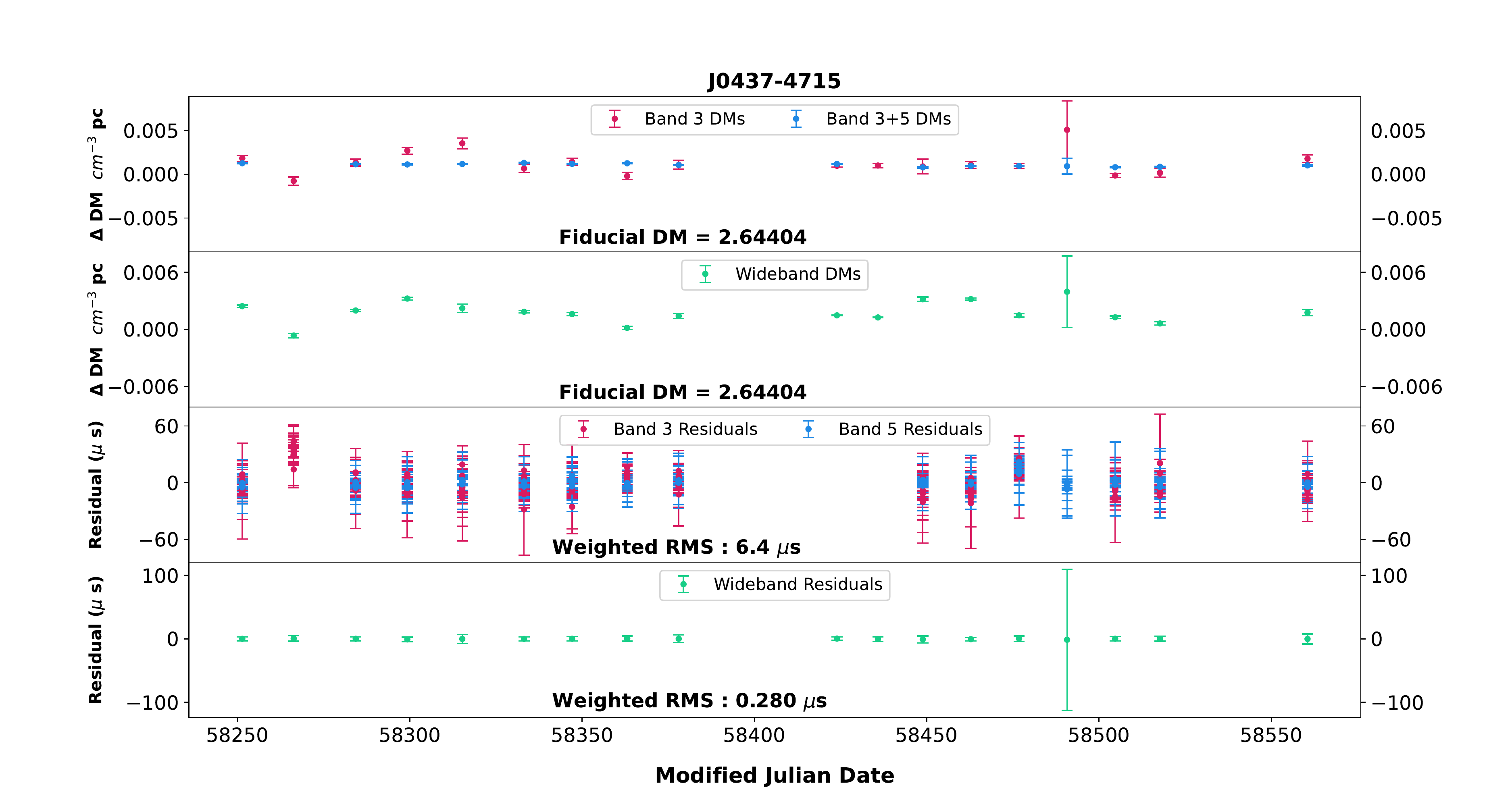}
\caption{Dispersion measure variations and timing residuals for J0437$-$4715 from 100 MHz observations. This pulsar was not observed during the later cycles with 200 MHz bandwidth. $\Delta$DMs (narrowband Band 3, narrowband Band 3+5, and wideband Band 3) represent the difference between estimated DMs and the fiducial DM (mentioned at the bottom of the corresponding panels). Narrowband and wideband timing residuals are shown in the two bottom panels (post-fit weighted RMS at the bottom of the respective panels).}
\label{fig:j0437}
\end{figure*}

\begin{figure*}[h!]
\includegraphics[width=0.8\linewidth]{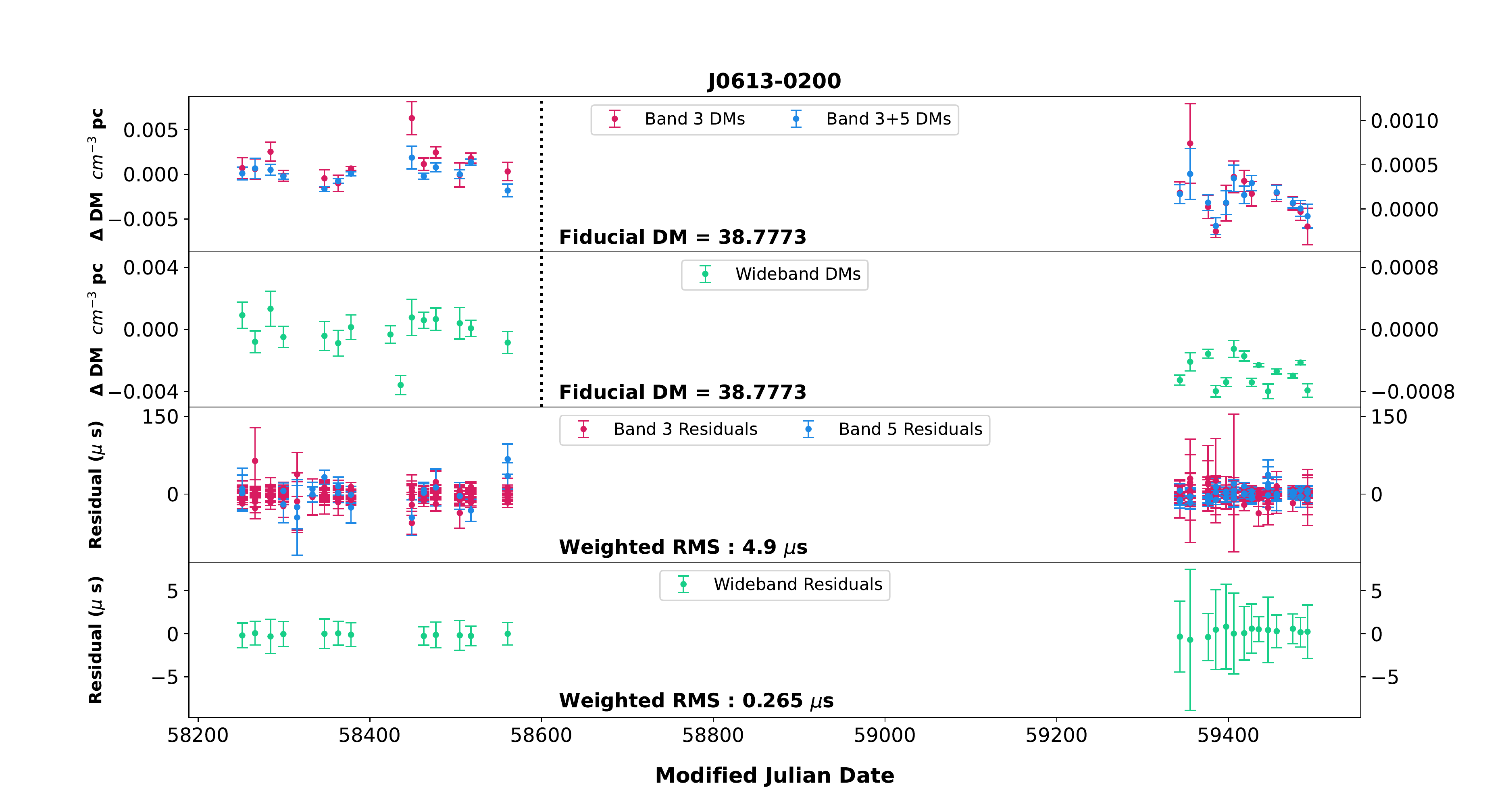}
\caption{Dispersion measure variations and timing residuals for J0613$-$0200. $\Delta$DMs (narrowband Band 3, narrowband Band 3+5, and wideband Band 3) for cycles 34-35 (100 MHZ bandwidth) are shown on the left y-axis and those for cycles 37-40 (200 MHZ bandwidth) are depicted on the right y-axis. The horizontal axis in the top two panels depicting $\Delta$DMs is split into two parts with a vertical dotted line along MJD 58600. The epochs on the left side of the dotted line represent 100 MHz bandwidth epochs while those on the right side represent 200 MHz bandwidth epochs. The y-axes for $\Delta$DMs on the left and right margins are scaled independently to make the $\Delta$DM variations visible. Narrowband and wideband timing residuals are shown in the two bottom panels. The fiducial DMs and post-fit weighted RMS residuals are mentioned at the bottom of the respective panels.}
\label{fig:j0613}
\end{figure*}

\begin{figure*}[h!]
\includegraphics[width=0.9\linewidth]{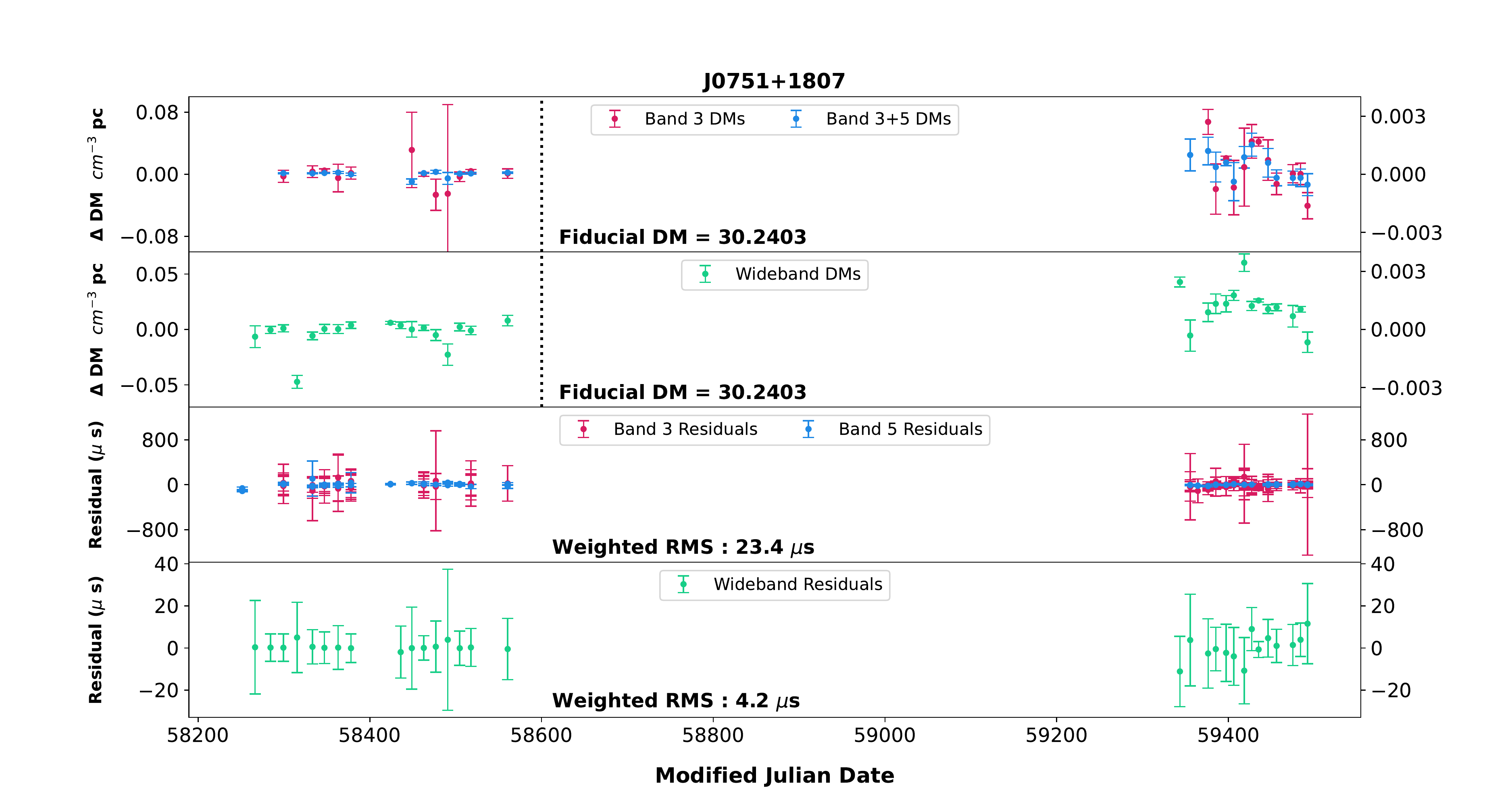}
\caption{Dispersion measure variations and timing residuals for J0751$+$1807. $\Delta$DMs (narrowband Band 3, narrowband Band 3+5 and wideband Band 3) for cycles 34-35 (100 MHZ bandwidth) are shown on the left y-axis and those for cycles 37-40 (200 MHZ bandwidth) are depicted on the right y-axis. The horizontal axis in the top two panels depicting $\Delta$DMs is split into two parts with a vertical dotted line along MJD 58600. The epochs on the left side of the dotted line represent 100 MHz bandwidth epochs while those on the right side represent 200 MHz bandwidth epochs. The y-axes for $\Delta$DMs on the left and right margins are scaled independently to make the $\Delta$DM variations visible. Narrowband and wideband timing residuals are shown in the two bottom panels. Fiducial DMs and post-fit weighted RMS residuals are mentioned at the bottom of the respective panels.}
\label{fig:0751}
\end{figure*}

\begin{figure*}[h!]
\includegraphics[width=0.9\linewidth]{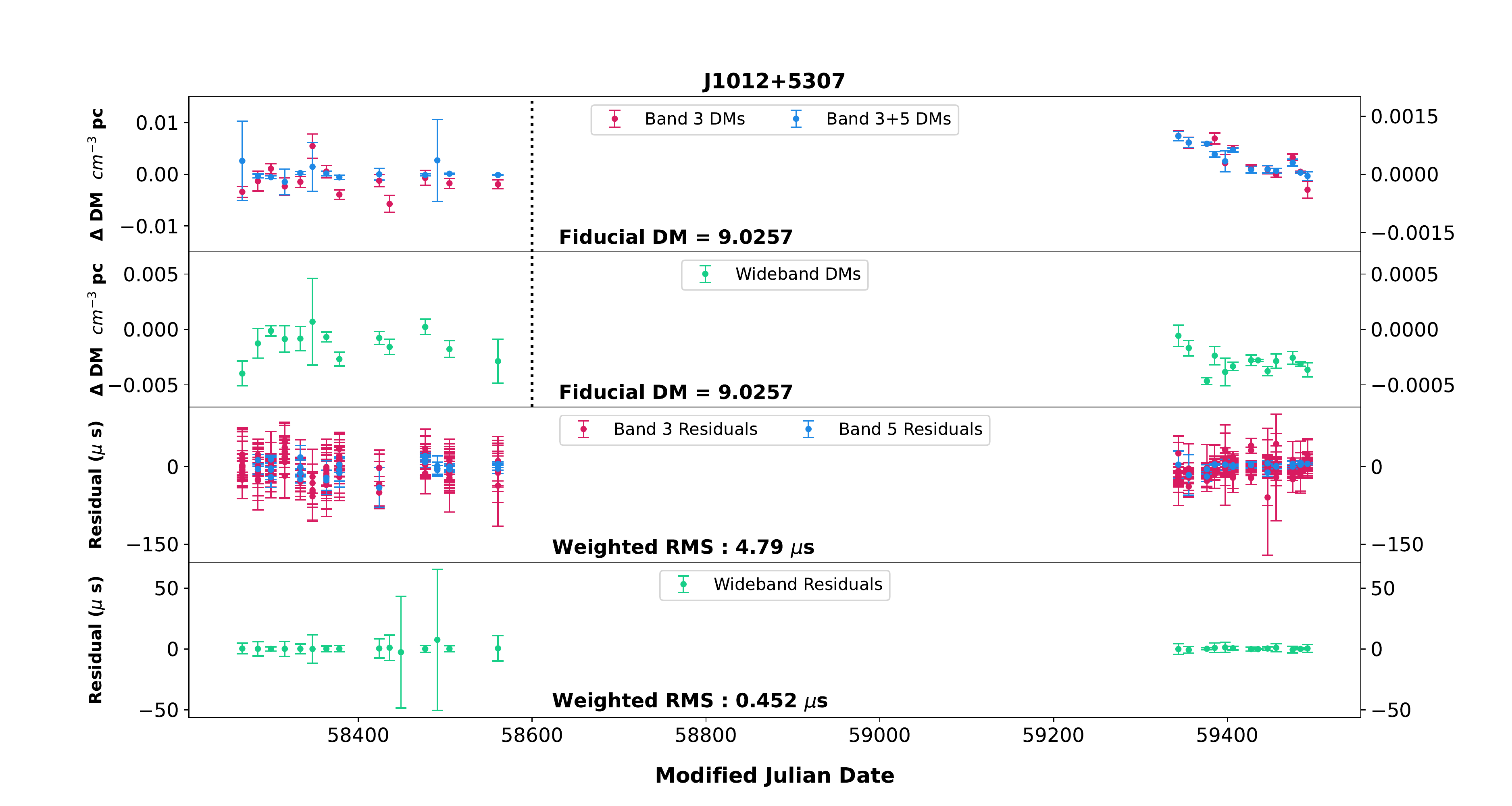}
\caption{Dispersion measure variations and timing residuals for J1012$+$5307. $\Delta$DMs (narrowband Band 3, narrowband Band 3+5, and wideband Band 3) for cycles 34-35 (100 MHZ bandwidth) are shown on the left y-axis and those for cycles 37-40 (200 MHZ bandwidth) are depicted on the right y-axis. The horizontal axis in the top two panels depicting $\Delta$DMs is split into two parts with a vertical dotted line along MJD 58600. The epochs on the left side of the dotted line represent 100 MHz bandwidth epochs while those on the right side represent 200 MHz bandwidth epochs. The y-axes for $\Delta$DMs on the left and right margins are scaled independently to make the $\Delta$DM variations visible. Narrowband and wideband timing residuals are shown in the two bottom panels. Fiducial DMs and post-fit weighted RMS residuals are mentioned at the bottom of the respective panels.}
\label{fig:j1012}
\end{figure*}

\begin{figure*}[h!]
\includegraphics[width=0.9\linewidth]{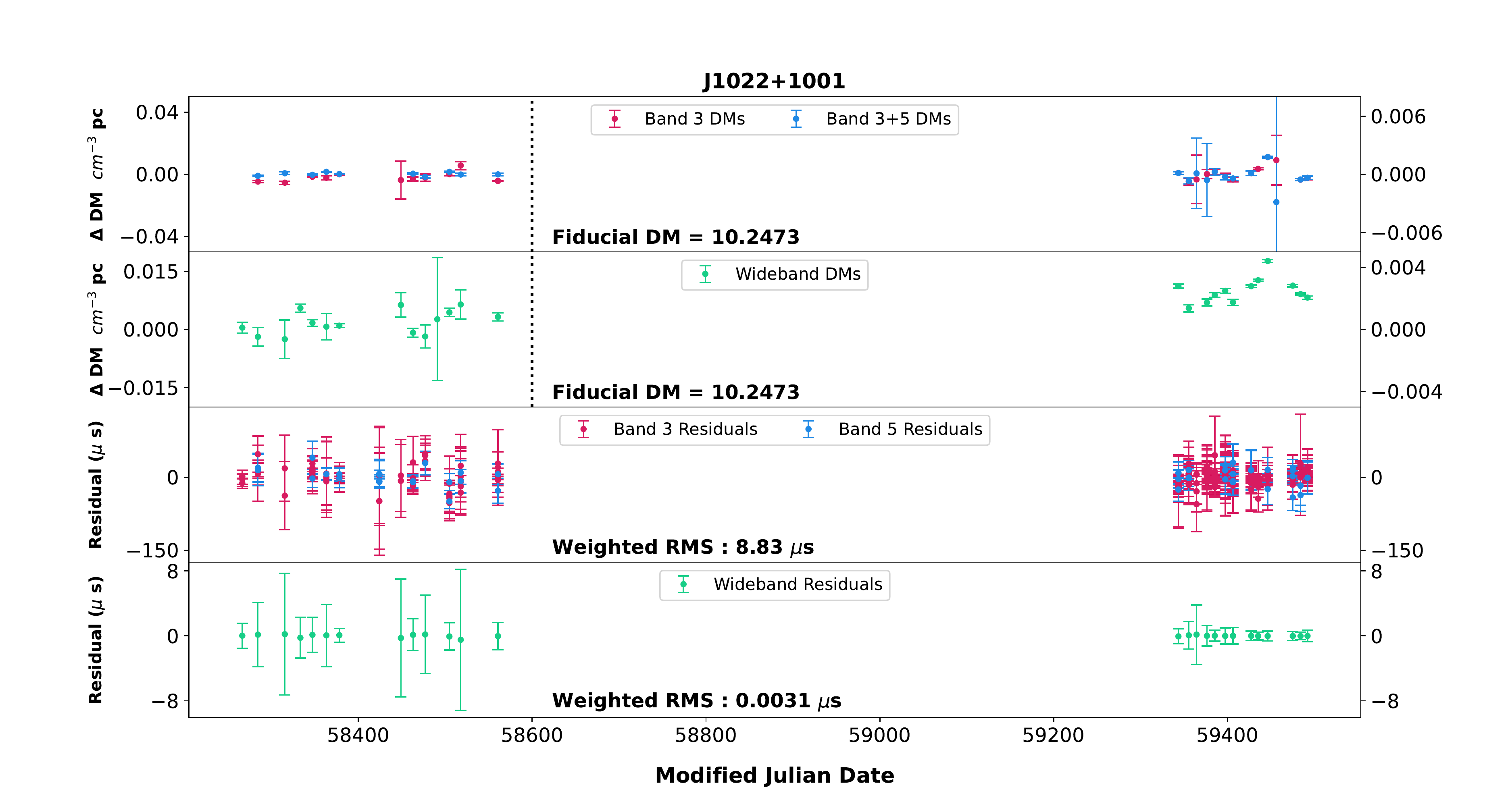}
\caption{Dispersion measure variations and timing residuals for J1022$+$1001. $\Delta$DMs (narrowband Band 3, narrowband Band 3+5, and wideband Band 3) for cycles 34-35 (100 MHZ bandwidth) are shown on the left y-axis and those for cycles 37-40 (200 MHZ bandwidth) are depicted on the right y-axis. The horizontal axis in the top two panels depicting $\Delta$DMs is split into two parts with a vertical dotted line along MJD 58600. The epochs on the left side of the dotted line represent 100 MHz bandwidth epochs while those on the right side represent 200 MHz bandwidth epochs. The y-axes for $\Delta$DMs on the left and right margins are scaled independently to make the $\Delta$DM variations visible. Narrowband and wideband timing residuals are shown in the two bottom panels. Fiducial DMs and post-fit weighted RMS residuals are mentioned at the bottom of the respective panels.}
\label{fig:j1022}
\end{figure*}

\begin{figure*}[h!]
\includegraphics[width=0.9\linewidth]{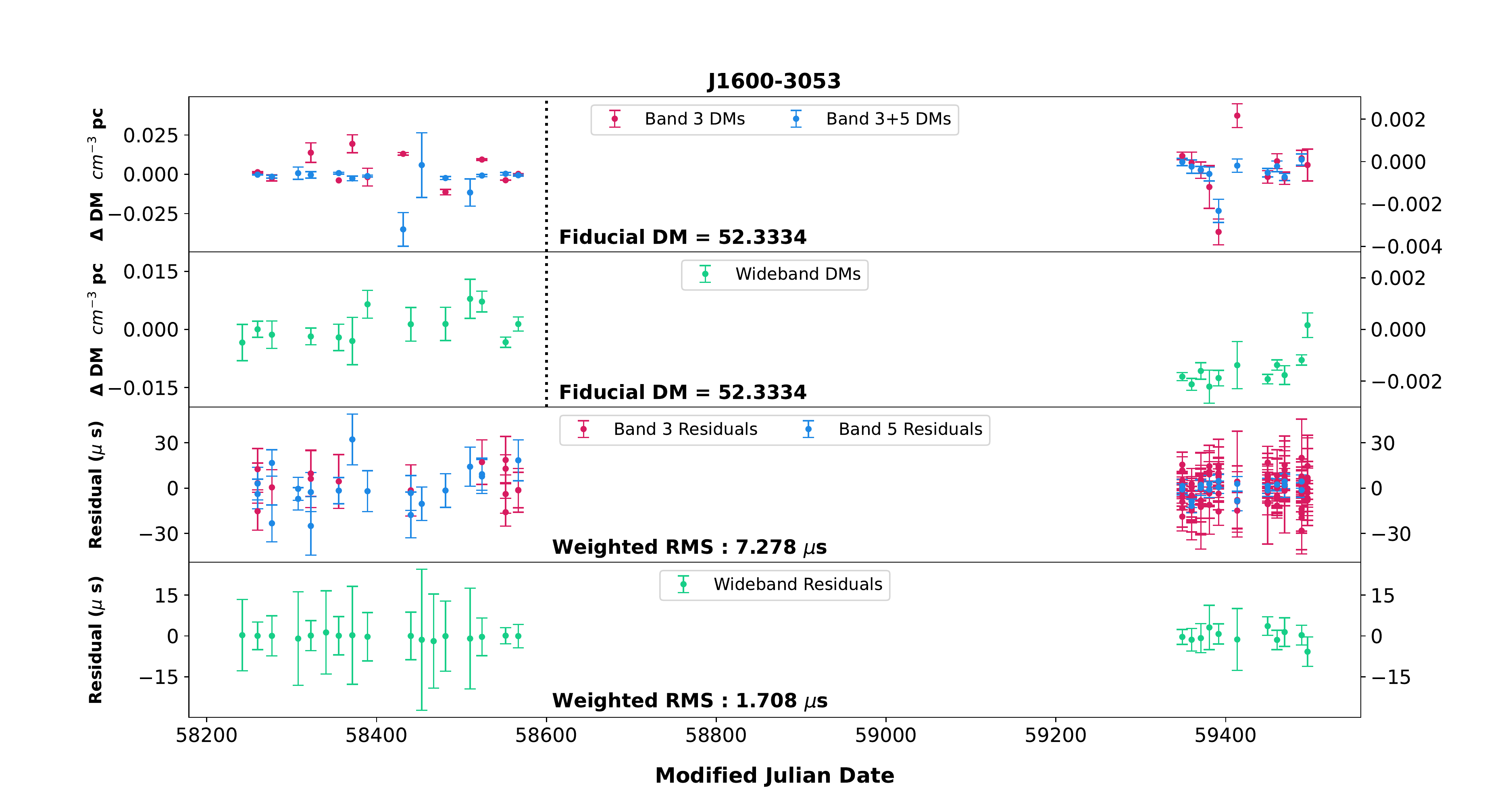}
\caption{Dispersion measure variations and timing residuals for J1600$-$3053. $\Delta$DMs (narrowband Band 3, narrowband Band 3+5, and wideband Band 3) for cycles 34-35 (100 MHZ bandwidth) are shown on the left y-axis and those for cycles 37-40 (200 MHZ bandwidth) are depicted on the right y-axis. The horizontal axis in the top two panels depicting $\Delta$DMs is split into two parts with a vertical dotted line along MJD 58600. The epochs on the left side of the dotted line represent 100 MHz bandwidth epochs while those on the right side represent 200 MHz bandwidth epochs. The y-axes for $\Delta$DMs on the left and right margins are scaled independently to make the $\Delta$DM variations visible. Narrowband and wideband timing residuals are shown in the two bottom panels. Fiducial DMs and post-fit weighted RMS residuals are mentioned at the bottom of the respective panels.}
\label{fig:j1600}
\end{figure*}

\begin{figure*}[h!]
\includegraphics[width=0.9\linewidth]{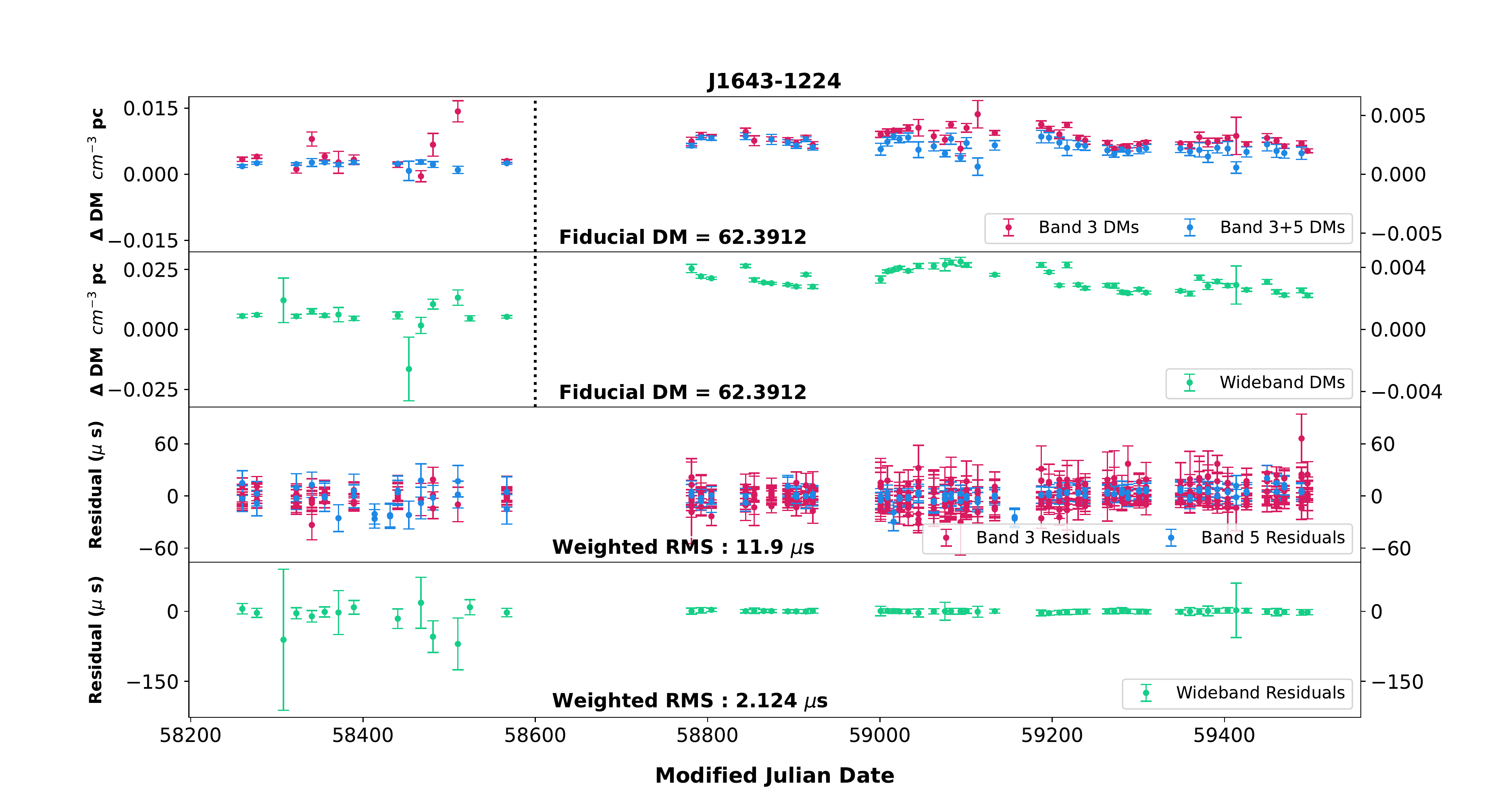}
\caption{Dispersion measure variations and timing residuals for J1643$-$1224. $\Delta$DMs (narrowband Band 3, narrowband Band 3+5, and wideband Band 3) for cycles 34-35 (100 MHZ bandwidth) are shown on the left y-axis and those for cycles 37-40 (200 MHZ bandwidth) are depicted on the right y-axis. The horizontal axis in the top two panels depicting $\Delta$DMs is split into two parts with a vertical dotted line along MJD 58600. The epochs on the left side of the dotted line represent 100 MHz bandwidth epochs while those on the right side represent 200 MHz bandwidth epochs. The y-axes for $\Delta$DMs on the left and right margins are scaled independently to make the $\Delta$DM variations visible. Narrowband and wideband timing residuals are shown in the two bottom panels. Fiducial DMs and post-fit weighted RMS residuals are mentioned at the bottom of the respective panels.}
\label{fig:j1643}
\end{figure*}

\begin{figure*}[h!]
\includegraphics[width=0.9\linewidth]{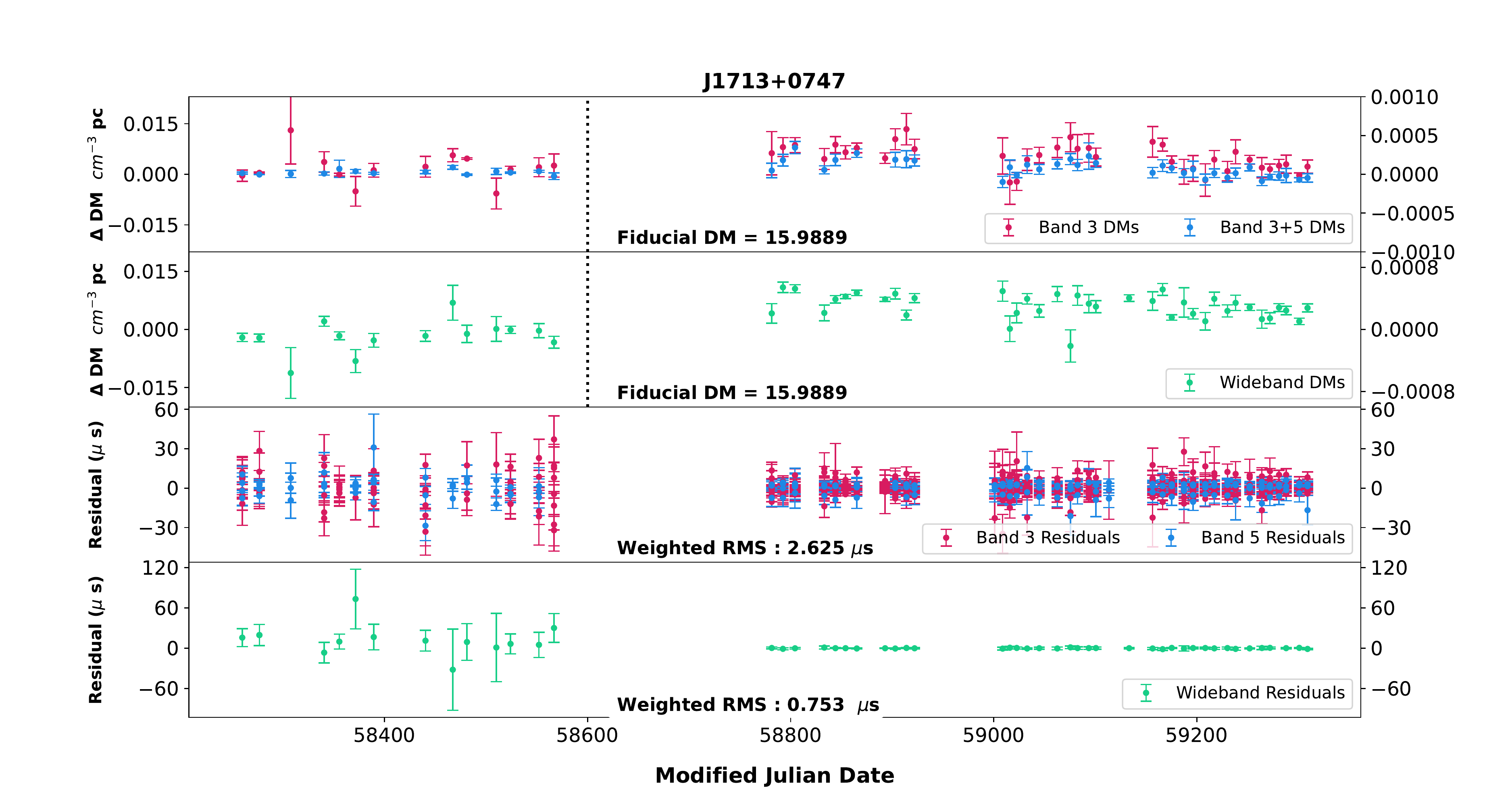}
\caption{Dispersion measure variations and timing residuals for J1713$+$0747. $\Delta$DMs (narrowband Band 3, narrowband Band 3+5, and wideband Band 3) for cycles 34-35 (100 MHZ bandwidth) are shown on the left y-axis and those for cycles 37-40 (200 MHZ bandwidth) are depicted on the right y-axis. The horizontal axis in the top two panels depicting $\Delta$DMs is split into two parts with a vertical dotted line along MJD 58600. The epochs on the left side of the dotted line represent 100 MHz bandwidth epochs while those on the right side represent 200 MHz bandwidth epochs. The y-axes for $\Delta$DMs on the left and right margins are scaled independently to make the $\Delta$DM variations visible. Narrowband and wideband timing residuals are shown in the two bottom panels. Fiducial DMs and post-fit weighted RMS residuals are mentioned at the bottom of the respective panels. Data beyond MJD 59320 is under investigation and has not been presented in the present release.}
\label{fig:j1713}
\end{figure*}

\begin{figure*}[h!]
\includegraphics[width=0.9\linewidth]{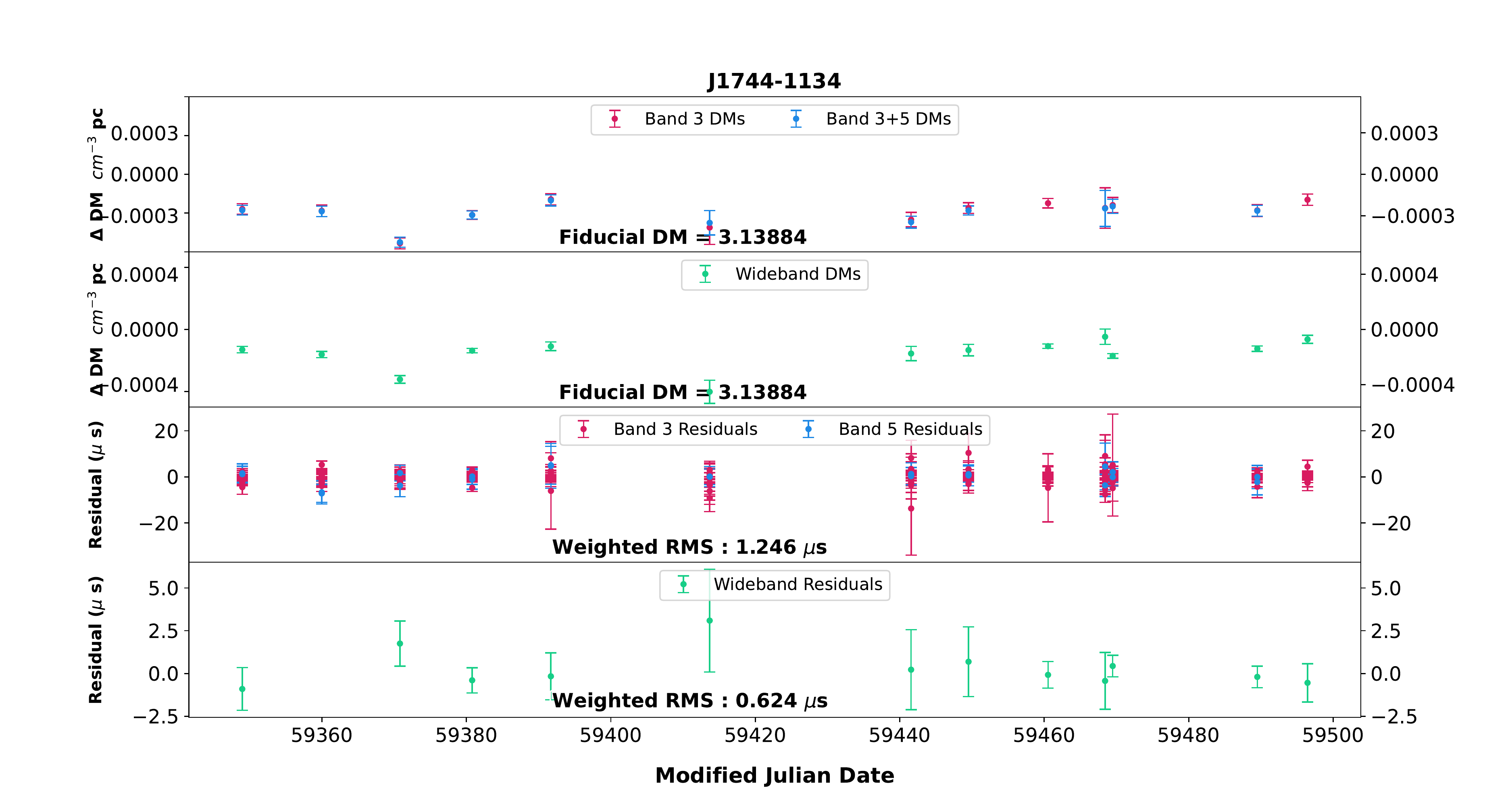}
\caption{Dispersion measure variations and timing residuals for J1744$-$1134. $\Delta$DMs (narrowband Band 3, narrowband Band 3+5, and wideband Band 3) from 200 MHZ bandwidth observations. This pulsar was not observed during the previous cycles. Narrowband and wideband timing residuals are shown in the two bottom panels. Fiducial DMs and post-fit weighted RMS residuals are mentioned at the bottom of the respective panels.}
\label{fig:j1744}
\end{figure*}

\begin{figure*}[h!]
\includegraphics[width=0.9\linewidth]{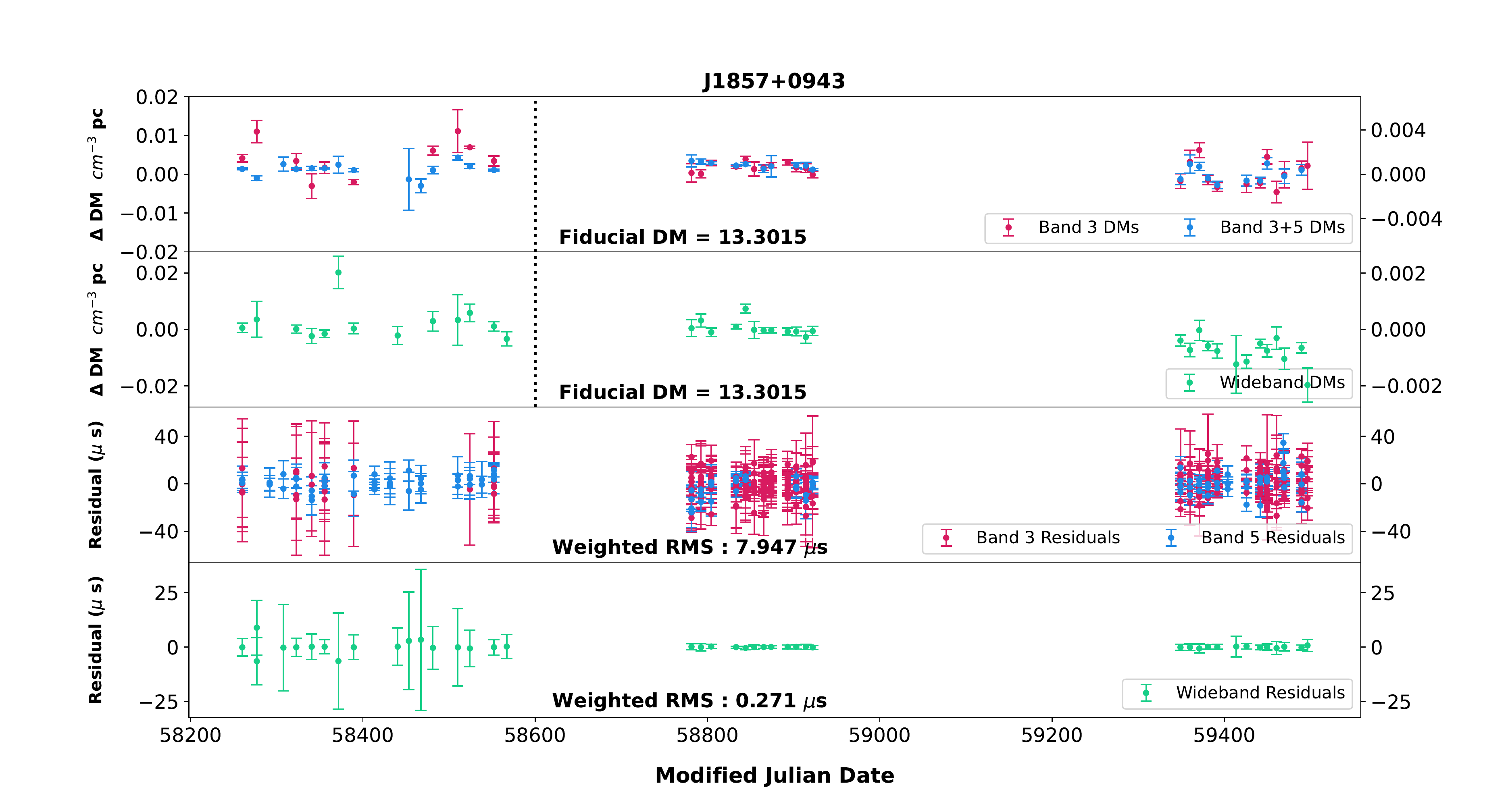}
\caption{Dispersion measure variations and timing residuals for J1857$+$0943. $\Delta$DMs (narrowband Band 3, narrowband Band 3+5, and wideband Band 3) for cycles 34-35 (100 MHZ bandwidth) are shown on the left y-axis and those for cycles 37-40 (200 MHZ bandwidth) are depicted on the right y-axis. The horizontal axis in the top two panels depicting $\Delta$DMs is split into two parts with a vertical dotted line along MJD 58600. The epochs on the left side of the dotted line represent 100 MHz bandwidth epochs while those on the right side represent 200 MHz bandwidth epochs. The y-axes for $\Delta$DMs on the left and right margins are scaled independently to make the $\Delta$DM variations visible. Narrowband and wideband timing residuals are shown in the two bottom panels. Fiducial DMs and post-fit weighted RMS residuals are mentioned at the bottom of the respective panels.}
\label{fig:j1857}
\end{figure*}

\begin{figure*}[h!]
\includegraphics[width=0.9\linewidth]{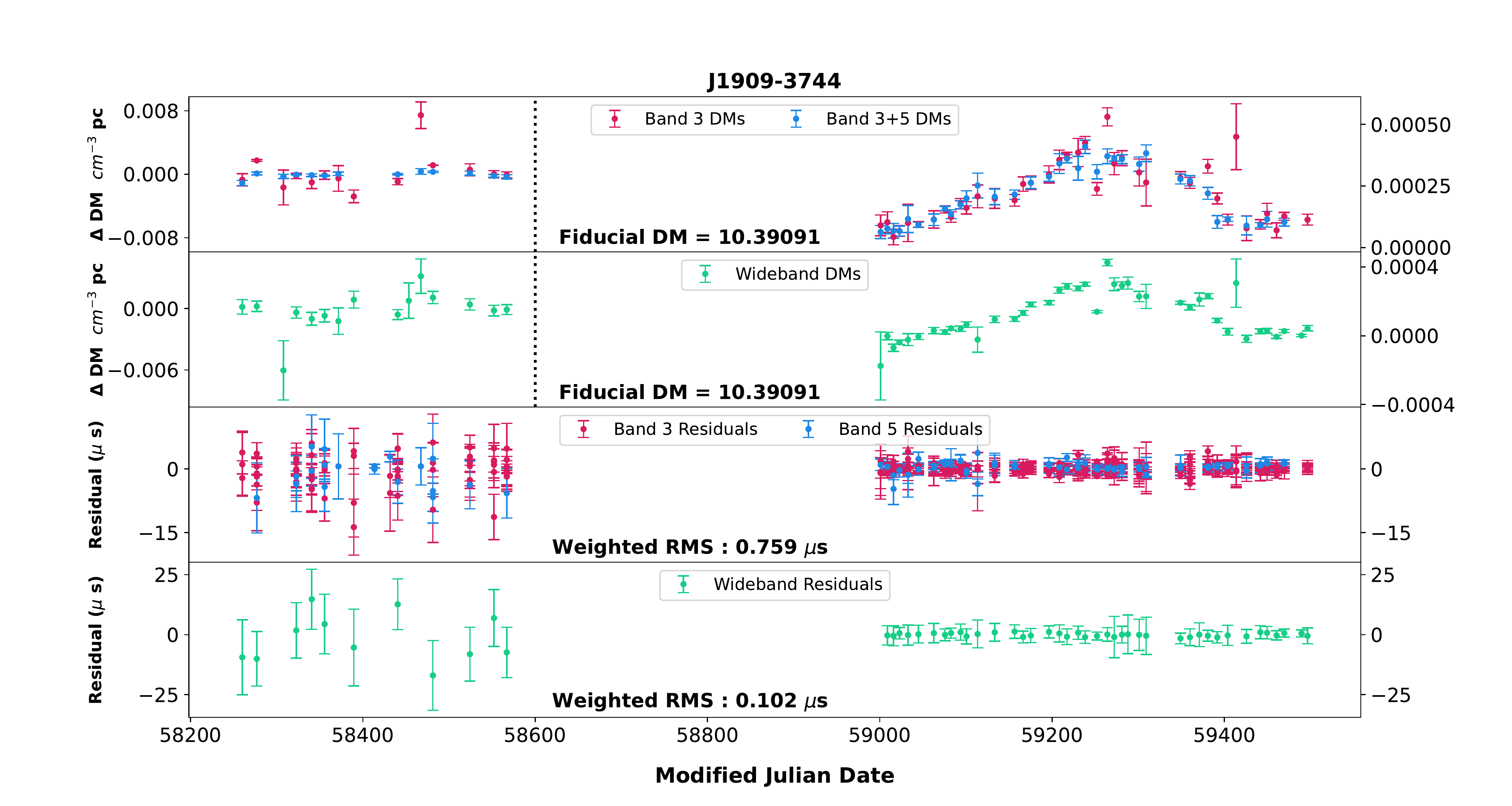}
\caption{Dispersion measure variations and timing residuals for J1909$-$3744. $\Delta$DMs (narrowband Band 3, narrowband Band 3+5, and wideband Band 3) for cycles 34-35 (100 MHZ bandwidth) are shown on the left y-axis and those for cycles 37-40 (200 MHZ bandwidth) are depicted on the right y-axis. The horizontal axis in the top two panels depicting $\Delta$DMs is split into two parts with a vertical dotted line along MJD 58600. The epochs on the left side of the dotted line represent 100 MHz bandwidth epochs while those on the right side represent 200 MHz bandwidth epochs. The y-axes for $\Delta$DMs on the left and right margins are scaled independently to make the $\Delta$DM variations visible. Narrowband and wideband timing residuals are shown in the two bottom panels. Fiducial DMs and post-fit weighted RMS residuals are mentioned at the bottom of the respective panels.}
\label{fig:j1909}
\end{figure*}

\begin{figure*}[h!]
\includegraphics[width=0.9\linewidth]{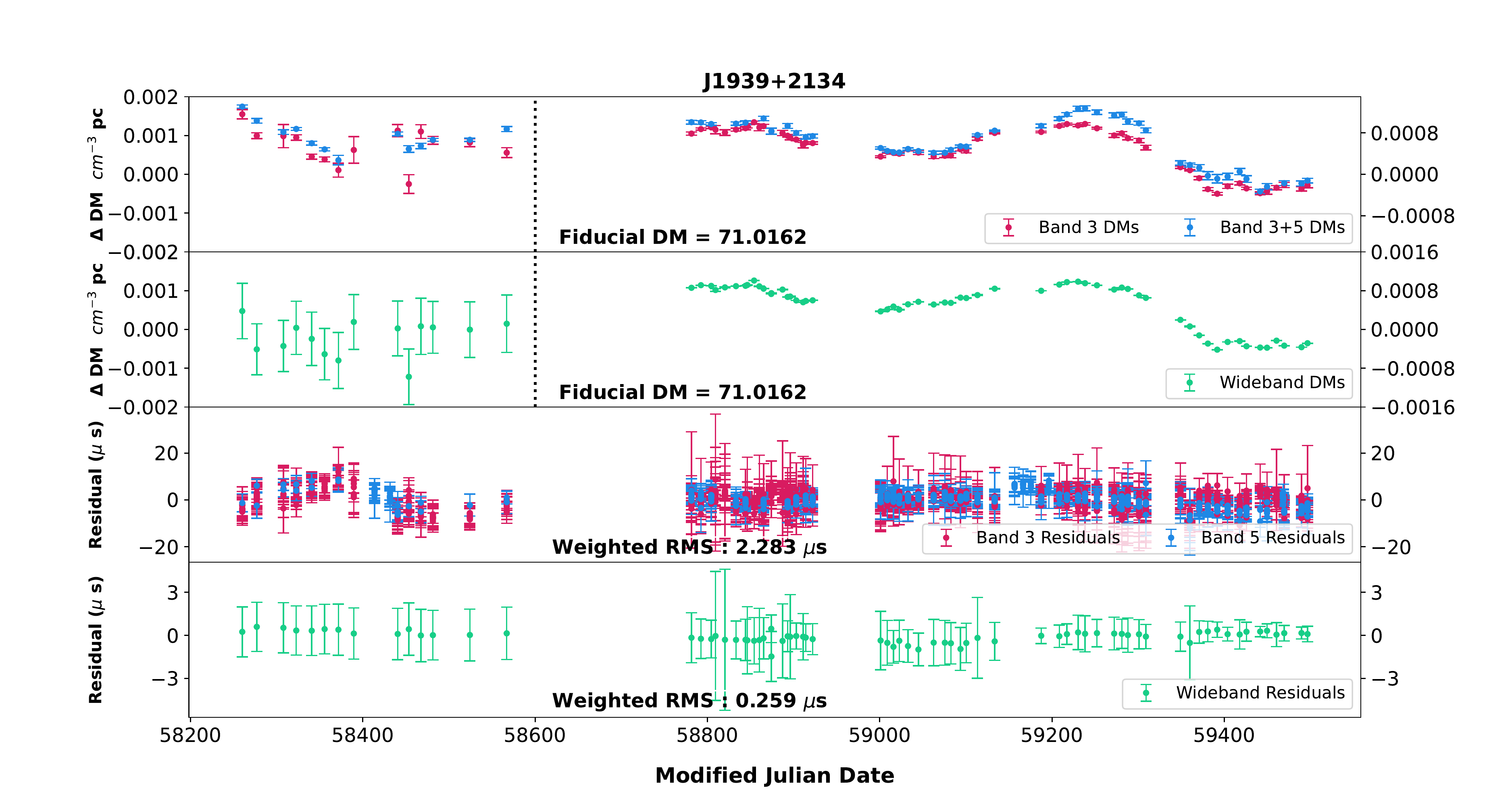}
\caption{Dispersion measure variations and timing residuals for J1939$+$2134. $\Delta$DMs (narrowband Band 3, narrowband Band 3+5, and wideband Band 3) for cycles 34-35 (100 MHZ bandwidth) are shown on the left y-axis and those for cycles 37-40 (200 MHZ bandwidth) are depicted on the right y-axis. The horizontal axis in the top two panels depicting $\Delta$DMs is split into two parts with a vertical dotted line along MJD 58600. The epochs on the left side of the dotted line represent 100 MHz bandwidth epochs while those on the right side represent 200 MHz bandwidth epochs. The y-axes for $\Delta$DMs on the left and right margins are scaled independently to make the $\Delta$DM variations visible. Narrowband and wideband timing residuals are shown in the two bottom panels. Fiducial DMs and post-fit weighted RMS residuals are mentioned at the bottom of the respective panels.}
\label{fig:j1939}
\end{figure*}

\begin{figure*}[h!]
\includegraphics[width=0.9\linewidth]{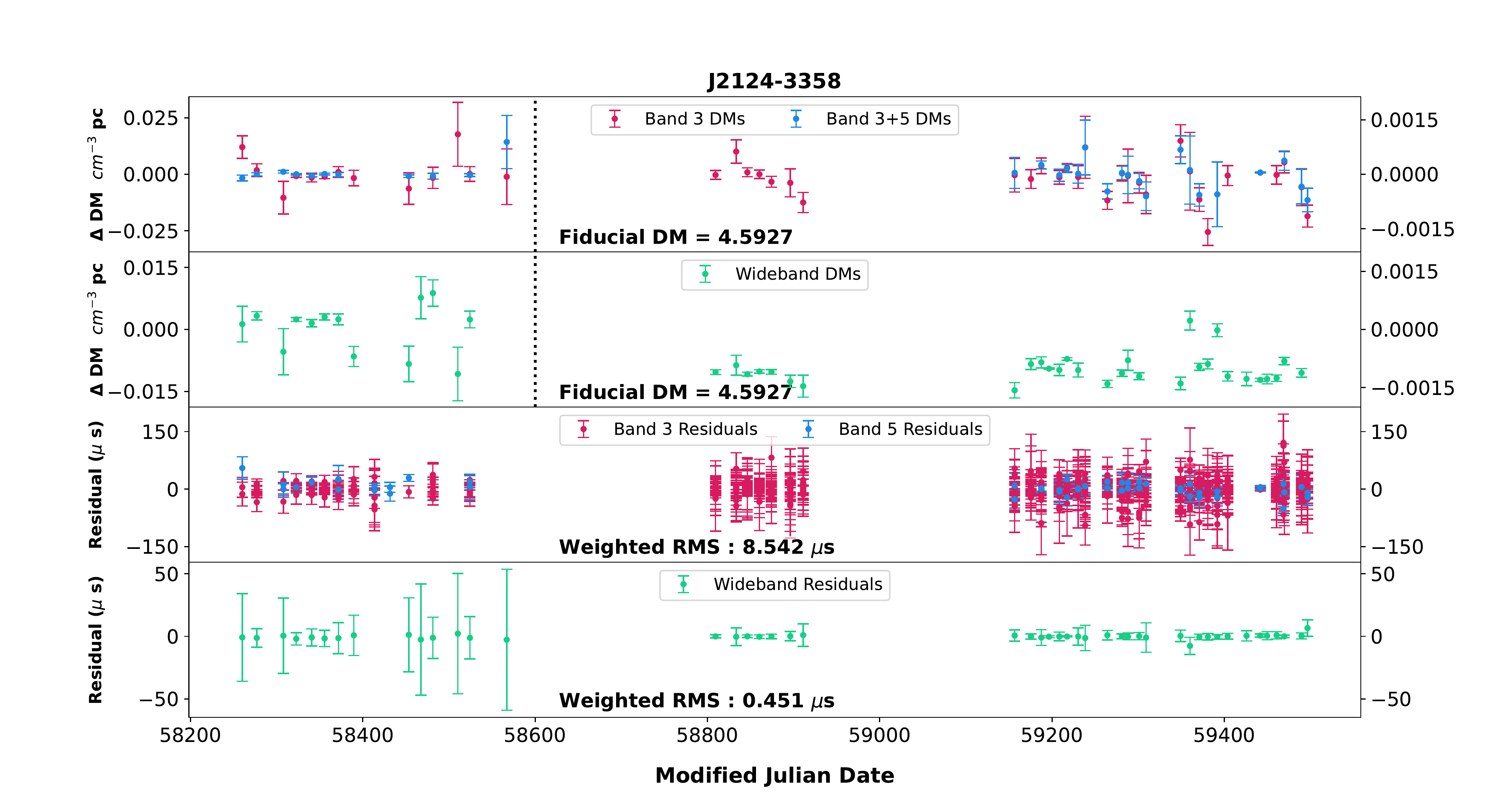}
\caption{Dispersion measure variations and timing residuals for J2124$-$3358. $\Delta$DMs (narrowband Band 3, narrowband Band 3+5 and wideband Band 3) for cycles 34-35 (100 MHZ bandwidth) are shown on the left y-axis and those for cycles 37-40 (200 MHZ bandwidth) are depicted on the right y-axis. The horizontal axis in the the top two panels depicting $\Delta$DMs is split into two parts with a vertical dotted line along MJD 58600. The epochs on left side of the dotted line represent 100 MHz bandwidth epochs while those on the right side represent 200 MHz bandwidth epochs. The y-axes for $\Delta$DMs on left and right margins are scaled independently to make the $\Delta$DM variations visible. Narrowband and wideband timing residuals are shown in the two bottom panels. Fiducial DMs and post-fit weighted RMS residuals are mentioned at the bottom of the respective panels.}
\label{fig:j2124}
\end{figure*}

\begin{figure*}[h!]
\includegraphics[width=0.9\linewidth]{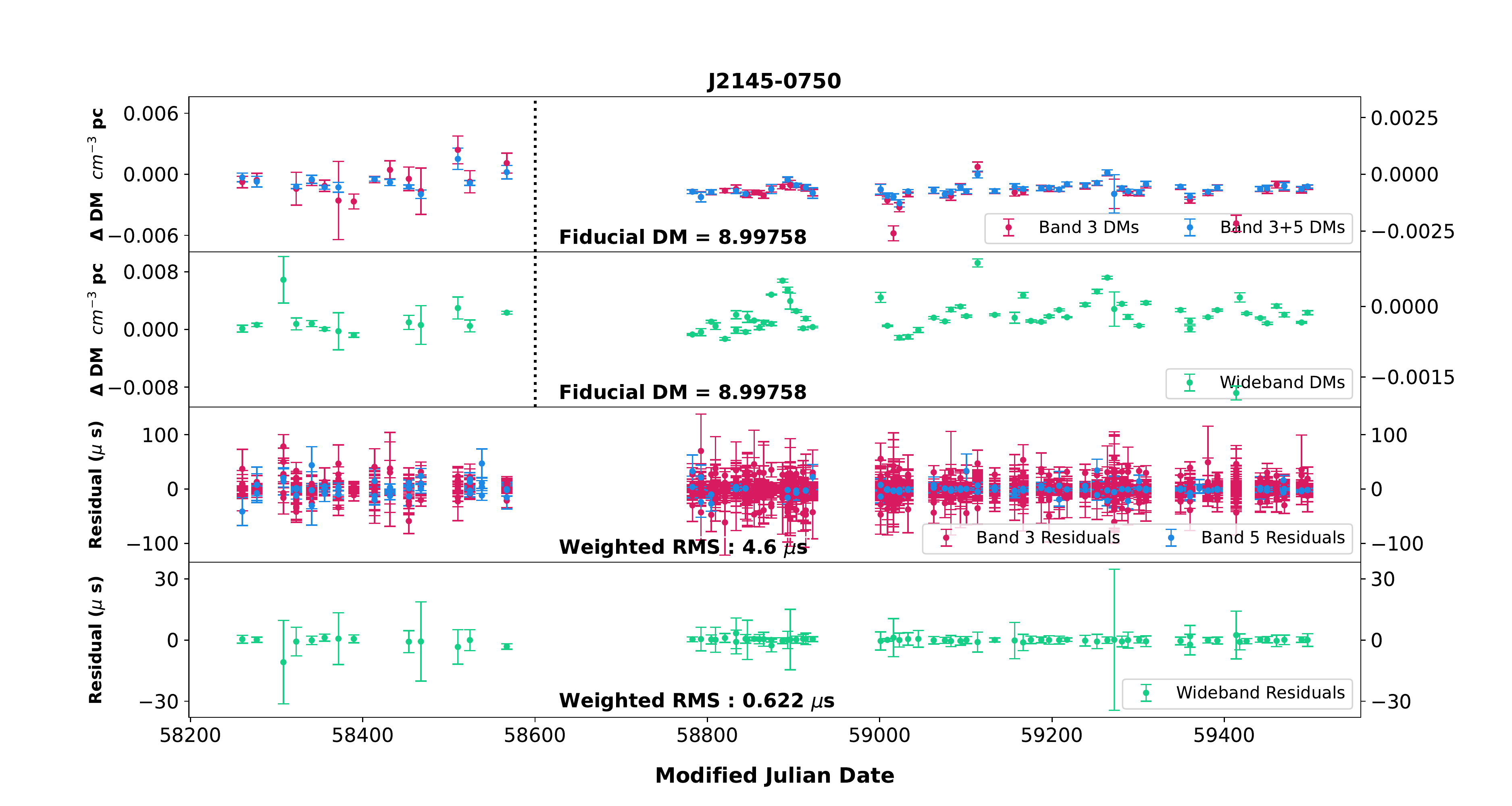}
\caption{Dispersion measure variations and timing residuals for J2145$-$0750. $\Delta$DMs (narrowband Band 3, narrowband Band 3+5, and wideband Band 3) for cycles 34-35 (100 MHZ bandwidth) are shown on the left y-axis and those for cycles 37-40 (200 MHZ bandwidth) are depicted on the right y-axis. The horizontal axis in the top two panels depicting $\Delta$DMs is split into two parts with a vertical dotted line along MJD 58600. The epochs on the left side of the dotted line represent 100 MHz bandwidth epochs while those on the right side represent 200 MHz bandwidth epochs. The y-axes for $\Delta$DMs on the left and right margins are scaled independently to make the $\Delta$DM variations visible. Narrowband and wideband timing residuals are shown in the two bottom panels. Fiducial DMs and post-fit weighted RMS residuals are mentioned at the bottom of the respective panels. The outliers in the 200 MHz NB Band 3 DMs (red points on the right side of the top panel) are presently under investigation.}
\label{fig:j2145}
\end{figure*}

\end{document}